\documentclass[a4paper,fleqn]{cas-sc}

\usepackage[numbers]{natbib}

\usepackage{color}
\usepackage{xcolor}
\usepackage{multirow}
\usepackage[table]{xcolor}

\usepackage{booktabs}
\usepackage{geometry}
\usepackage{lineno}
\usepackage{amssymb}

\usepackage{graphicx}
\usepackage{caption}
\usepackage{tabularx}
\usepackage{ltablex}
\usepackage{amsfonts}
\keepXColumns

\usepackage{amsmath} 

\usepackage{algorithm}
\usepackage{algorithmic}

\usepackage{tikz}
\usetikzlibrary{arrows.meta, positioning, shapes.geometric, fit}

\usepackage{subcaption}

\def\tsc#1{\csdef{#1}{\textsc{\lowercase{#1}}\xspace}}
\tsc{WGM}
\tsc{QE}
\tsc{EP}
\tsc{PMS}
\tsc{BEC}
\tsc{DE}

\begin{document}
\let\WriteBookmarks\relax
\def\floatpagepagefraction{1}
\def\textpagefraction{.001}
\shorttitle{Recommender for Model Selection}
\shortauthors{S. Hänsch et~al.}

\title [mode = title]{Hybrid Cold-Start Recommender System for Closure Model Selection in Multiphase Flow Simulations}                      
\tnotemark[1]

\tnotetext[1]{This work was supported by funding from the European Union for the project CoSAS, co-financed from tax revenues on the basis of the budget adopted by the Saxon State Parliament; and by the Helmholtz Initiative "HPC Gateway" within the pilot project "Center for Industrial Multiphase Flow Competence".}

\author[1]{S. Hänsch}[type=editor,
                        auid=000,bioid=1,
                        orcid=0000-0003-1296-5566]
\cormark[1]
\ead{s.haensch@hzdr.de}

\credit{Data curation, Writing - Original draft preparation, Conceptualization, Methodology, Investigation}

\affiliation[1]{organization={Institute of Fluid Dynamics, Helmholtz-Zentrum Dresden-Rossendorf e.V.},
                addressline={Bautzner Landstraße 400}, 
                postcode={01328}, 
                city={Dresden},
                country={Germany}}

\author[2]{A. Sajdoková}[orcid=0009-0005-0552-7455]
\ead{sajdokova@gmail.com}
\credit{Conceptualization, Methodology, Visualization}

\affiliation[2]{organization={Charles University},
                addressline={Ovocný trh 560/5}, 
                postcode={116 36}, 
                city={Prague 1},
                country={Czech Republic}}

\author[3]{A. Rębowski}[orcid=0009-0001-5083-3882]
\ead{rabau.andrei@gmail.com}
\credit{Investigation, Methodology, Visualization}
\affiliation[3]{organization={AGH University of Kraków},
                addressline={al. Adama Mickiewicza 30}, 
                city={Kraków},
                postcode={30-059}, 
                country={Poland}}

\author[4]{F. Miškařík}[orcid=0009-0005-3924-5843]
\ead{miskafil@fit.cvut.cz}
\credit{Methodology, Visualization, Writing – review \& editing}
\affiliation[4]{organization={Faculty of Information Technology, Czech Technical University in Prague},
                addressline={Thákurova 7}, 
                postcode={16000}, 
                city={Prague 6 - Dejvice},
                country={Czech Republic}}

\author[5]{K. Ramakrishna}[orcid=0000-0003-4211-2484]
\ead{k.ramakrishna@hzdr.de}
\credit{Data curation, Writing – review \& editing}
\affiliation[5]{organization={Information Services and Computing, Helmholtz-Zentrum Dresden-Rossendorf e.V.},
                addressline={Bautzner Landstraße 400}, 
                postcode={01328}, 
                city={Dresden},
                country={Germany}}

\author[1]{F. Schlegel}[orcid=0000-0003-3824-9568]
\ead{f.schlegel@hzdr.de}
\credit{Software,  Writing – review \& editing}

\author[4]{V. Rybář}[orcid=0000-0003-0552-3997]
\ead{vojtech.rybar@fit.cvut.cz}
\credit{Methodology, Writing – review \& editing, Supervision}

\author[4]{R. Alves}[orcid=0000-0001-7458-5281]
\ead{rodrigo.alves@fit.cvut.cz}
\credit{Conceptualization, Methodology, Writing – Original draft preparation, Supervision}

\author[4]{P. Kordík}[orcid=0000-0003-1433-0089]
\ead{pavel.kordik@fit.cvut.cz}
\credit{Conceptualization, Methodology, Writing – review \& editing, Supervision}

\cortext[cor1]{Corresponding author}

\begin{abstract}
Selecting appropriate physical models is a critical yet difficult step in many areas of computational science and engineering. In multiphase Computational Fluid Dynamics (CFD), practitioners must choose among numerous closure model combinations whose performance varies strongly across flow conditions. Sub-optimal choices can lead to inaccurate predictions, simulation failures, and wasted computational resources, making model selection a prime candidate for data-driven decision support.
This work formulates closure model selection as a cold-start recommender system problem in a high-cost scientific domain. We propose a hybrid recommendation framework that combines (i) metadata-driven case similarity and (ii) collaborative inference via matrix completion. The approach enables case-specific model recommendations for entirely new CFD cases using their descriptive features, while leveraging historical simulation results from similar cases.
The methodology is evaluated on 13,600 simulations across 136 validation cases and 100 model combinations. A nested cross-validation protocol with experiment-level holdout is employed to rigorously assess generalisation to unseen flow scenarios under varying levels of data sparsity. Recommendation quality is measured using ranking-based metrics and a domain-specific regret measure capturing performance loss relative to the per-case optimum.
Results show that the proposed hybrid recommender consistently outperforms popularity-based and expert-designed reference models and reduces regret across the investigated sparsities. These findings demonstrate that recommender system methodology can effectively support complex scientific decision-making tasks characterised by expensive evaluations, structured metadata, and limited prior observations. Beyond CFD, the work highlights the broader potential of recommendation frameworks to improve efficiency, reliability, and resource usage in computational science and engineering workflows.
\end{abstract}



\maketitle

\section{Introduction}

In times when the need for sustainability and the complexity of processes is constantly increasing, multiphase Computational Fluid Dynamics (CFD) plays a crucial role in the digitalisation of process optimisation. In particular, Eulerian-Eulerian multiphase CFD provides a continuum-based framework that enables the detailed description of complex, interacting multiphase flows at scales relevant to industrial components \citep{Yeoh2019, Garcia-Villalba2025}. While the framework allows coarse computational grids, fast results at moderate costs and the use of complex geometries, a high number of physical phenomena are not resolved directly but instead must be represented by closure models. For bubbly flows these phenomena include interfacial forces, turbulence and bubble-bubble interactions, while reactive multiphase flows additionally require modelling of heat and mass transfer as well as chemical reactions \citep{Ishii2011}. As the complexity of the flows increases, so does the number of physical phenomena that need to be modelled using suitable closure models. 

Driven by the increasing use of CFD across diverse application domains, the last few decades have seen the development of an extensive portfolio of closure models addressing a wide range of multiphase flow scenarios. While this constantly expanding model portfolio reflects the scientific progress, it poses a practical challenge when setting up CFD simulations: selecting the appropriate closure models for a given flow scenario. The selection of closure models best suited for a particular multiphase flow problem is acknowledged to be a key issue to make reliable and accurate CFD predictions \citep{Gray2021, Besagni2023, Adzaklo2025}. While recent advances in Large Language Models (LLMs) have begun to automate aspects of CFD simulation setup and execution \citep{Fan2026, Pandey2025, Dong2025}, existing LLM-based agents primarily focus on workflow orchestration rather than on the selection of appropriate physical models. Their success rate diminishes in the context of multiphase flows, where the underlying physics is highly complex and governed by numerous tightly coupled closure models \citep{Fan2026}. The intricate model interdependencies present a combinatorial selection challenge that extends beyond pattern recognition in configuration workflows. In practice, closure model selection in multiphase CFD continues to rely heavily on expert knowledge, as well as substantial time and computational resources for systematic model assessment and sensitivity analysis.

In an attempt to address the model selection dilemma in bubbly flow research, significant efforts were made over the years to develop universally applicable models. By prioritising mechanistic models grounded in local flow conditions rather than empirical ones derived from isolated experiments, these works aim to establish closure model sets that remain valid across a wide range of flow conditions \citep{Lucas2016a, Rzehak2017, Liao2020, Colombo2021, Garcia-Villalba2025}. Despite the efforts, no consensus has yet emerged in the multiphase community regarding a ``one-size-fits-all'' closure model set capable of delivering reliable predictions for arbitrary flow configurations.

Rather than pursuing universal closure formulations, an alternative solution strategy is to embrace the diversity of closure models while facilitating their appropriate selection by filtering and customising the selection to the different flow scenarios. Information filtering, and recommender systems in particular, are successfully applied in other domains where decision making is hindered by an overload of possible options \citep{Misir2017, schedl2019deep, alves2024regionalization}. As noted by \citet{Huang2023} and \citet{Haensch2025} closure model selection can be formalised as a collaborative filtering problem, by considering that a CFD validation case ``prefers'' the closure models that achieve a better performance in terms of the accuracy of results. However, closure model recommendation differs fundamentally from conventional recommender system applications: In contrast to typical user–item settings, it involves extremely high evaluation costs, relies on structured scientific metadata rather than behavioural feedback, and carries safety- and reliability-relevant consequences. Closure model recommendation therefore represents a distinct class of Artificial Intelligence (AI)-supported decision-making problems in scientific computing. Unlike consumer recommendation tasks, errors in this context can propagate into physically misleading predictions and costly design decisions, making risk-aware evaluation essential. The work of \citet{Haensch2025} proposes storing and organising CFD results in shared validation databases to preserve the collective knowledge about which closure models and parameters were successful in the past. Matrix completion methods were shown to be efficient in predicting the best closure models even when based on extremely sparse performance data for a few case-model interactions.

This paper builds upon the idea of addressing the closure model selection problem from an information filtering perspective. Similarly to previous approaches, a performance matrix created via batch-processing of a validation database of CFD cases is used to train and evaluate a recommender system model for the problem of closure model selection for multiphase flows. In contrast, this work explicitly incorporates CFD case features to characterise validation experiments and identify physically similar cases, enabling recommendations for entirely new, unseen CFD scenarios without requiring prior simulations. The novelty of this work lies not in the individual components of the proposed recommender, but in their principled integration for high-cost scientific decision-making, combined with a realistic cold-start evaluation protocol at the level of entire experiments, which reflects practical deployment conditions in CFD workflows.

In Section 2 we describe the data set used in this study, including the investigated CFD validation cases, their associated feature labels, and the explored closure model combinations, which represent a subset of the vast closure model problem space. The batch-processing workflow and subsequent extraction of performance values for case-model interactions used to construct the performance matrix are outlined.

Section 3 introduces recommender system models in general while making the analogy to our closure model selection task. We then describe the proposed hybrid recommender model based on matrix completion and feature-based case similarity. 

In Section 4 we present the evaluation methodology adopted for the recommender model. We define our item relevance, the evaluation metrics used, and describe the cross-validation and testing procedures. 

Section 5 discusses results obtained with the recommender model for the individual test experiments. We then compare the performance of the proposed recommender model against both a conventional expert-designed reference model and a data-driven popularity baseline. Exemplary validation plots are added to illustrate cases with low and high potential for recommendation improvements.

Section 6 summarises the main findings of this study, examines their implications for closure model selection in multiphase CFD as a representative application of scientific computing, and identifies promising directions for future research.

\section{Dataset}
\label{sec:data}

\subsection{CFD validation cases}

This work benefits from a shared validation database of CFD simulation setups for use with the OpenFOAM Foundation software \citep{HZDRcases}. This collaborative database currently offers around 700 multiphase flow cases from 31 different experiments, including bubbly flow, flotation and hybrid cases covering both dispersed and segregated gas structures. CFD setups of new validation cases from previous research studies can be stored in this publicly available repository and easily shared with the scientific community.

For the creation of our performance matrix from this database a selection of 17 different bubbly flow experiments was made as listed in Table \ref{tab:caseList}. Note that each experiment consists of 1-39 sub-cases with unique flow parameters, summing up to a total of 136 CFD cases. The simulation setups and references are available with the repository \citep{HZDRcases}. Each of the individual CFD cases is labelled with a total of 27 categorical and continuous case features, as described below. 

\begin{table}[htbp]
\centering
\captionsetup{width=1.0\linewidth}
\footnotesize
\setlength{\tabcolsep}{3pt} 
\caption{List of the explored experiments with corresponding experiment and case identifiers (IDs) for reference, and their parameter ranges for the list of continuous features: bubble diameter $d_{Air}$, superficial gas ($J_G$) and liquid velocities ($J_L$), pressure $p$, temperature $T$ and the characteristic size $D$ of the flow configuration. Detailed descriptions of these experiments and their simulation setup can be found in the public repository of \citet{HZDRcases}.}
\begin{tabular}{lllllllll}
\toprule
\textbf{Exp.} & \textbf{Case}  & \textbf{Experiment} & \textbf{$d_{Air}$} & \textbf{$J_G$} & \textbf{$J_L$} & \textbf{$p$} & \textbf{$T$} & \textbf{$D$}\\
\textbf{ID} & \textbf{IDs}  &  & \textbf{[mm]} & \textbf{[m/s]} & \textbf{[m/s]} & \textbf{[kPa]} & \textbf{[K]} & \textbf{[m]}\\
\midrule
1  & 1-6   & \citet{Sommer2023}        & 2.45--4.47 & 0.006               & 0                     & 101.3              & 298.15 & 0.1\\
2  & 7-10  & \citet{NeumannKipping2020}      & 3.73--5.43 & 0.0368              & 0.405--1.017          & 366.7              & 303.15 & 0.053\\
3  & 11-28 & \citet{Ziegenhein2019}  & 4.88--6.94 & $2.96\times10^{-3}$  & 0                     & 101.3              & 298.15 & 0.1125\\
4  & 29-32 & \citet{Kim2016}      & 2.2--3.7   & $[4,38]\times10^{-4}$& 0.0187--0.045         & 101.3              & 298.15 & 0.04\\
5  & 33-36 & \citet{Hosokawa2013} & 2.62--3.59 & $[4,30]\times10^{-4}$& 0.045                & 101.3              & 298.15 & 0.02 \\
6  & 37-41 & \citet{Lucas2010}    & 3.88--4.37 & 0.0062--0.0151      & 0.641--1.017          & 165.3              & 303.23 & 0.1953\\
7  & 42-47 & \citet{Mudde2009}    & 4.25       & 0.015--0.049        & 0                     & 101.3              & 298.15 & 0.15 \\
8  & 48-51 & \citet{Hosokawa2009} & 3.21--4.25 & 0.018--0.036        & 0.50--1.0             & 101.3              & 298.15 & 0.025\\
9  & 52-55 & \citet{Shawkat2008}  & 3.20--5.00 & 0.015--0.1          & 0.45--0.68            & 101.3              & 297.65 & 0.2 \\
10 & 56-95 & \citet{Lucas2005}    & 2.37--13.36& 0.004--0.534        & 0.102--4.047          & 107.8--134.7       & 302.68--304.35 & 0.0512\\
11 & 96-101& \citet{Hibiki2001}   & 2.6--3.4   & 0.027--0.471        & 0.49--2.01            & 101.3              & 293.15 & 0.0508\\
12 & 102   & \citet{Deen2001}     & 4.0       & 0.0049             & 0                     & 101.3              & 298.15 & 0.15\\
13 & 103   & \citet{Pfleger1999}  & 3.0       & $1.3\times10^{-3}$  & 0                     & 101.3              & 298.15 & 0.2\\
14 & 104-107& \citet{Liu1998}     & 2.94--4.22 & 0.1197--0.2196      & 0.51--1.05            & 101.3              & 299.15 & 0.0572\\
15 & 108-109& \citet{LiuBankoff1993}     & 2.3--3.2   & 0.027--0.112        & 1.087                 & 101.3              & 283.15 & 0.038\\
16 & 110-133& \citet{Wang1987}    & 2.33--3.41 & 0.06--0.42          & 0.35--1.02            & 101.3              & 284.3  & 0.05715\\
17 & 134-136& \citet{SunFaeth1986} & 1.0       & $[2,8]\times10^{-5}$& 0                     & 101.3              & 293.15 & 0.534\\
\bottomrule
\label{tab:caseList}
\end{tabular}
\end{table}

\subsubsection{Categorical features}

The current data set includes 17 categorical features (mostly binary) that label the characteristics of both the experimental and the simulation setup of the CFD cases used here. An overview of the whole list of possible unique feature entries for our data set with descriptions is given in Table \ref{tab:catFeatures}. Presumably, these features are known a priori without running a CFD simulation as they describe e.g. an expected flow regime and Re number classification, or void fraction estimates (above/below 0.1). Note that the feature list of the entire validation data base, to be found as keyword list in \cite{HZDRcases}, is more extensive as it covers a more diverse collection of multiphase flow cases, while we here focus on a selection of bubbly flow cases only.

\begin{table}[htbp]
\centering
\captionsetup{width=0.95\linewidth}
\footnotesize
\caption{List of categorical case features with possible entries covered by the explored data set.}
\renewcommand{\arraystretch}{1.15}
\begin{tabular}{c l p{0.15\linewidth} p{0.5\linewidth}}
\toprule
\textbf{Setup} & \textbf{Category} & \textbf{Options} & \textbf{Description}\\
\midrule
\multirow{22}{*}{\rotatebox[origin=c]{90}{\textbf{Experimental}}} &
\textbf{Configuration} &
bubble column\newline
pipe\newline
turbulent jet
& Type of experimental flow facility and geometric arrangement.\\
\addlinespace[0.4em]  
& \textbf{Contamination} &
filtered\newline
tap\newline
contaminated
& Level of liquid purity, ranging from filtered to intentionally contaminated.\\
\addlinespace[0.4em]
& \textbf{Cross section shape} &
circular\newline
rectangular
& Shape of the flow cross section perpendicular to the main flow direction.\\
\addlinespace[0.4em]
& \textbf{Gas–liquid flow regime} &
bubbly\newline
finely-dispersed bubbly\newline
slug
& Characteristic gas–liquid flow pattern and interfacial structure.\\
\addlinespace[0.4em]
& \textbf{Gas injection} &
bottom injection\newline
side injection
& Location where gas is introduced into the liquid flow.\\
\addlinespace[0.4em]
& \textbf{Gas void fraction level} &
low\newline
high
& Average gas volume fraction in the measurement region with low-alpha referring to levels below and high-alpha above 0.1.\\
\addlinespace[0.4em]
& \textbf{Internals} &
ring\newline
baffle
& Internal structures placed inside the channel to obstruct or guide the flow.\\
\addlinespace[0.4em]
& \textbf{Orientation} &
vertical upward\newline
vertical downward
& Direction of the main flow relative to gravity.\\
\addlinespace[0.4em]
& \textbf{Reynolds number} &
high\newline
low
& Flow regime based on the Reynolds number (laminar or turbulent).\\
\midrule

\multirow{17}{*}{\rotatebox[origin=c]{90}{\textbf{Simulation}}} &
\textbf{Bubble size} &
monodisperse\newline
polydisperse
& Whether a single bubble size or a size distribution is modeled.\\
\addlinespace[0.4em]
& \textbf{Bubble size changes} &
coalescence/breakup\newline
density change\newline
fixed
& Mechanisms allowing bubble sizes to evolve or remain fixed.\\
\addlinespace[0.4em]
& \textbf{Development} &
fully developed\newline
developing
& Whether streamwise flow development is resolved in the simulation.\\
\addlinespace[0.4em]
& \textbf{Gas phases} &
one gas\newline
multiple gas
& Number of distinct gas phases included in the model.\\
\addlinespace[0.4em]
& \textbf{Geometry} &
2D–wedge\newline
3D
& Dimensionality and geometric representation of the computational domain.\\
\addlinespace[0.4em]
& \textbf{Inlet condition} &
two-phase inlet\newline
gas volume source
& Boundary condition prescribing how gas and liquid enter the domain.\\
\addlinespace[0.4em]
& \textbf{Outlet condition} &
degassing\newline
pressure
& Boundary condition used to remove phases or control pressure.\\
\addlinespace[0.4em]
& \textbf{State} &
steady state\newline
transient
& Whether the simulation is time-independent or time-resolved.\\

\bottomrule

\label{tab:catFeatures}
\end{tabular}
\end{table}

\subsubsection{Continuous features}

In addition to the categorical features, we add six continuous features that are extracted directly from the simulation setups in a consistent manner. These features do not require CFD simulation, as they are input parameters specified during simulation setup. The continuous features comprise of the bubble diameter $d_{Air}$, computed as the Sauter mean diameter for polydisperse cases, the superficial gas ($J_G$) and liquid ($J_L$) velocities, pressure $p$, Temperature $T$ and the characteristic size of the respective flow configuration $D$ (e.g. the radius for pipes or the column width for bubble columns). Table \ref{tab:caseList} gives an overview of the feature parameter range for the individual CFD experiments explored in this work.

The selection of continuous features is restricted to the ones that were available for every single CFD case belonging to the data set. More continuous features are available from experimental data or reference simulation results, but cause the issue of missing entries for certain cases. 

\subsection{Closure models}

Eulerian-Eulerian multiphase CFD is based on an averaged set of continuity and momentum equations for each phase, gas and liquid, containing terms for certain physical phenomena that require modelling. The momentum transfer between gas and liquid is typically described as interfacial forces, i.e. drag, virtual mass, turbulent dispersion and wall lubrication forces. 

In the literature there is an abundance of model options for each of these phenomena leading to a combinatorial explosion of possible model configurations \citep{Haensch2025}. Considering only model options available with the code repository used for this study \citep{HZDRcode}, the total number of possible combinations exceeds half a million. The fact that many more models are available in the literature, and that inherent model parameters have not yet been considered, demonstrates the enormous task of model selection faced by CFD practitioners today. 

Given the extensive model catalogue, for this study we limit the model selection space to a smaller sub-domain. We keep most of the aforementioned models fixed and focus on the lateral force combinations available with the code repository used \citep{HZDRcode}, as illustrated in Table \ref{tab:itemPortfolio}. The possible options for shear lift, turbulent dispersion and wall lubrication forces, leaves us with a total of 100 closure model combinations with the model parameters kept at their default values. 

Due to the tightly coupled nature of interfacial forces, turbulence modelling, and bubble–bubble interactions, the findings of this study cannot be directly extrapolated to the full space of available closure models. Accordingly, the data and results presented are not intended to support physical interpretation or general conclusions regarding individual model fidelity. Rather, they are used exclusively to investigate and demonstrate the potential of recommender system methodologies as a decision-support tool for closure model selection.

\begin{table}[ht]
\centering
\footnotesize
\caption{Overview of the investigated closure model combinations for bubbly flow as available from \citet{HZDRcode} with the fixed default model parameters used in this study. Varied model options for shear lift, turbulent dispersion and wall lubrication forces are highlighted in gray. A reference model designed by researchers for general applicability \citep{Garcia-Villalba2025} is highlighted in bold. More detailed descriptions of the listed models can be found in the public repository of \citet{HZDRcode}.}
\setlength{\tabcolsep}{4pt}
\renewcommand{\arraystretch}{1.15}

\begin{tabularx}{\linewidth}{ll l X}
\toprule
\multicolumn{2}{l}{\textbf{Required Closures}} 
& \textbf{Investigated Model Options} 
& \textbf{Default Model Parameters} \\
\midrule

\multirow{16}{*}{\textbf{Interfacial Forces}} 
& Drag 
& \citet{IshiiZuber1979} 
& -- \\

& Virtual mass 
& \citet{Crowe2011} 
& $C_{\mathrm{VM}} = 0.5$ \\

\addlinespace[0.4em]
& \multirow{5}{*}{Shear lift}
& \cellcolor{gray!12}None & -- \\
& & \cellcolor{gray!12}\citet{Legendre1998} & -- \\
& & \cellcolor{gray!12}\citet{Moraga1999} & -- \\
& & \cellcolor{gray!12}\citet{Tomiyama2002b} & -- \\
& & \cellcolor{gray!12}\textbf{\citet{Hessenkemper2021}} & -- \\

\addlinespace[0.4em]
& \multirow{4}{*}{Turbulent dispersion}
& \cellcolor{gray!12}None & -- \\
& & \cellcolor{gray!12}\citet{Gosman1992} & -- \\
& & \cellcolor{gray!12}\citet{LopezDeBertodano1992} & -- \\
& & \cellcolor{gray!12}\textbf{\citet{Burns2004}} & $\sigma = 0.9$ \\

\addlinespace[0.4em]
& \multirow{5}{*}{Wall lubrication}
& \cellcolor{gray!12}None & -- \\
& & \cellcolor{gray!12}\citet{Antal1991} 
& $C_{W1}=-0.01,\; C_{W2}=0.05$ \\
& & \cellcolor{gray!12}\citet{Tomiyama1998b} 
& $C_{WD}=0.15$ \\
& & \cellcolor{gray!12}\textbf{\citet{Hosokawa2002}} 
& -- \\
& & \cellcolor{gray!12}\citet{Frank2005} 
& $C_{WC}=10,\; C_{WD}=6.8,\; p=1.7$ \\

\midrule

\multirow{2}{*}{\textbf{Turbulence}}
& Shear-induced
& \citet{Menter2003}
& $\alpha_{K1}=0.85,\ \alpha_{K2}=1.0,\ \alpha_{\omega1}=0.5,\ \alpha_{\omega2}=0.856,$ \\
& & &
$\beta_1=0.075,\ \beta_2=0.0828,\ \beta^\ast=0.09,$ \\
& & &
$\gamma_1=\tfrac{5}{9},\ \gamma_2=0.44,\ a_1=0.31,\ b_1=1.0,\ c_1=10.0$ \\

\addlinespace[0.4em]
& Bubble-induced
& \citet{Ma2017}
& -- \\

\midrule

\multirow{2}{*}{\textbf{Polydispersity}}
& Coalescence
& \citet{Liao2015}
& $C_{\mathrm{Eff}}=2.5$; 
$C_{\mathrm{Turb}}=C_{\mathrm{Buoy}}=C_{\mathrm{Shear}}$ \\
& & &
$=C_{\mathrm{Eddy}}=C_{\mathrm{Wake}}=1.0$ \\

\addlinespace[0.4em]
& Breakup
& \citet{Liao2015}
& $B_{\mathrm{Turb}}=B_{\mathrm{Shear}}=B_{\mathrm{Eddy}}=1.0$; $B_{\mathrm{Fric}}=0.25$ \\

\bottomrule
\end{tabularx}
\label{tab:itemPortfolio}
\end{table}

\subsection{Performances for case-model interactions}

Previous works \citep{Haensch2021, Lehnigk2023} introduced a fully automated workflow for the large-scale pre-processing, execution, and post-processing of CFD simulations using the \textit{Snakemake} workflow management system \citep{snakemakeLibrary}. These tools were employed in the present work to generate simulation results in the form of validation plots for the selection of 136 CFD cases and 100 closure model combinations. The resulting outputs, amounting to tens of thousands of validation plots, were systematically evaluated and compared using a fuzzy-logic-based performance assessment.

The fuzzy logic controller \citep{Haensch2021} aggregates error metrics for the discrepancy between simulation and validation data into a single performance score between 0 and 1. This score provides a compact measure of agreement between simulation predictions and experimental reference data. Detailed descriptions of the batch-processing pipeline, the individual error metrics, and the fuzzy-logic inference system are available in the cited literature and in the publicly released evaluation scripts \citep{HZDRpython}.

Note that the performance metric used in this study aggregates multiple physical error measures into a single scalar score. While this enables systematic ranking and comparison of model combinations, it necessarily compresses multidimensional simulation quality into a one-dimensional representation. In practice, different applications may prioritise different physical quantities, suggesting the potential for multi-objective or preference-aware extensions of the framework in the future.

Leveraging this automated evaluation methodology enables the present study to cover an unusually broad spectrum of the closure model domain. Amongst the many investigated closure model combinations are case-model interactions that lead to numerical instabilities, manifested either as diverging simulations or as spatially oscillatory (``staggering" or zigzag) solution profiles. Because such artifacts primarily reflect deficiencies in numerical stability rather than physical model fidelity, they are deliberately not captured by the fuzzy-logic error metrics, which are designed to quantify the physical accuracy. These failure modes are therefore detected and treated separately.

Staggering patterns are identified by analysing the cell-centre solution fields for repeated sign changes in the spatial gradients. Specifically, a solution is flagged as unstable if at least five consecutive directional changes exceed 1\% of the global range of the corresponding field, indicating non-physical oscillations. The detection function used with default parameters is publicly available alongside the other evaluation scripts \citep{HZDRpython}. Simulations that either crash or exhibit such oscillatory behaviour are assigned a performance value of zero. Although this conflates physical inaccuracy and numerical instability, it allows the present analysis to proceed with a single performance matrix; a more refined separation into accuracy and stability metrics is deferred to future work.

\begin{figure}
  \centering
  \includegraphics[width=1.0\textwidth]{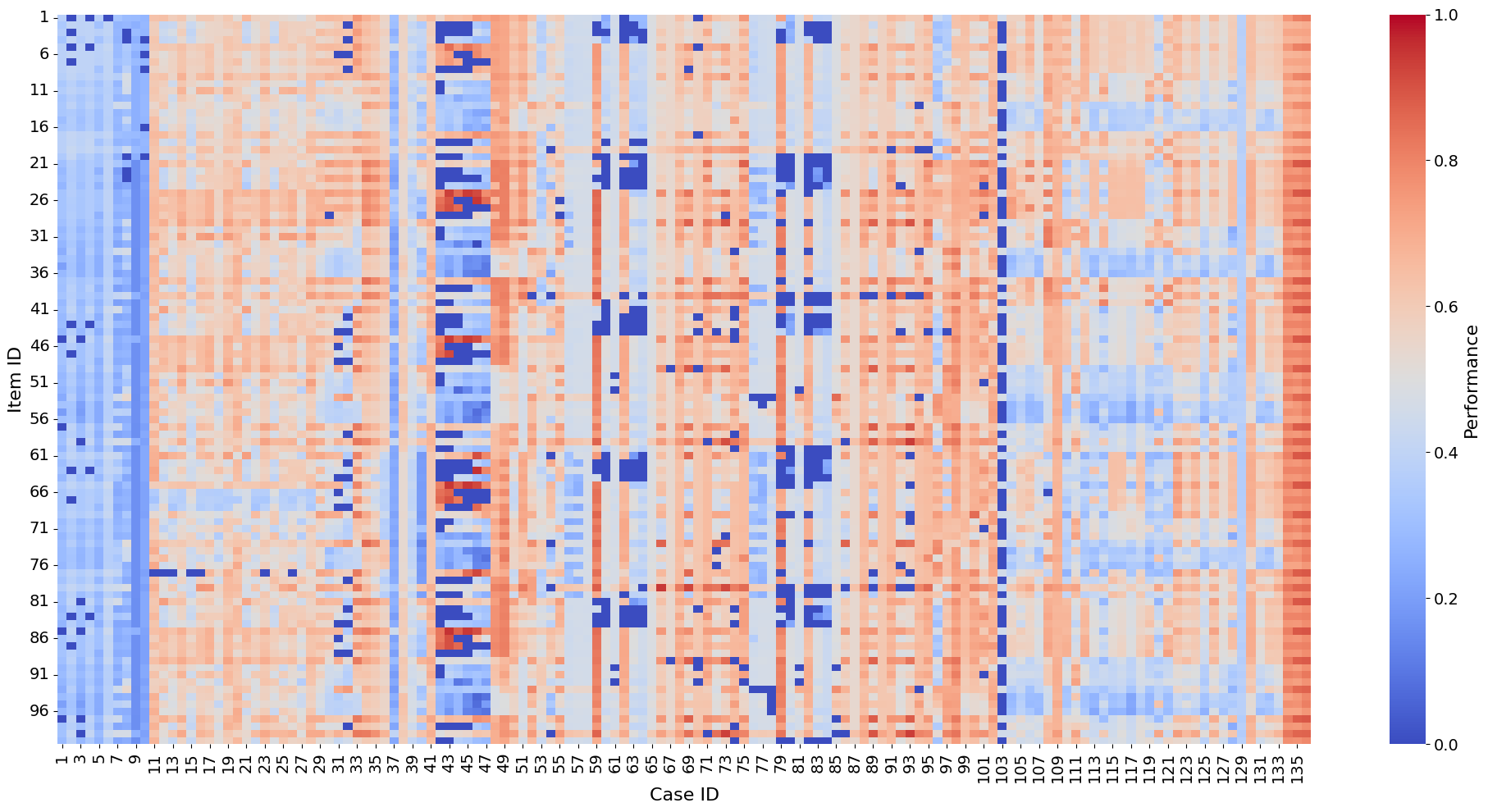}
  \caption{The ground-truth performance matrix constructed from 100 items evaluated for 136 CFD cases, with the performance values indicated by colour coding. Case-item interactions that lead to numerical instabilities present themselves in dark blue (performance value of zero).}
  \label{fig:performanceMatrix}
\end{figure}

Figure \ref{fig:performanceMatrix} presents the resulting ground-truth performance matrix comprising 13,600 performance values for all investigated case-model interactions. Entries corresponding to unstable or failed simulations appear as dark blue (zero performance). Column-wise clustered performance values reveal the inherent bias present in the different CFD cases: some consistently achieve good agreement with experiments, whereas others remain difficult to predict regardless of the model selected. Row-wise, across a CFD experiment and its sub-cases, a varying degree of sensitivity to the choice of model is observed. In practical applications, only a sparse subset of this ground-truth matrix will be observable, and future deployments will span a much larger portion of the closure-model problem space.

\subsection{Case representation}

\begin{figure}
  \centering

  \begin{subfigure}[t]{0.48\textwidth}
    \centering
    \includegraphics[width=\linewidth]{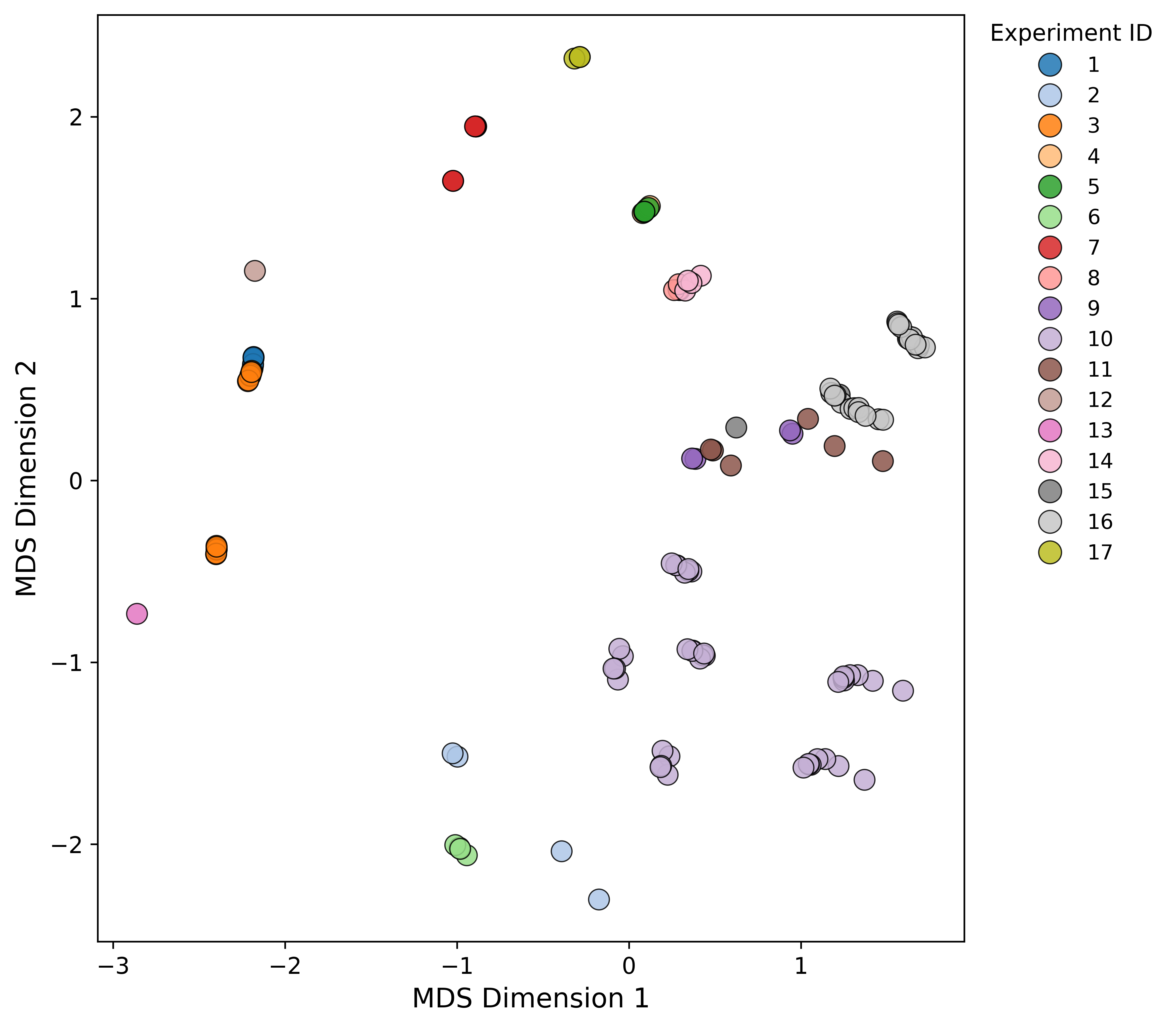}
    \caption{MDS projection of the CFD cases using the feature space.}
    \label{fig:mds_initial_features}
  \end{subfigure}
  \hfill
  \begin{subfigure}[t]{0.48\textwidth}
    \centering
    \includegraphics[width=\linewidth]{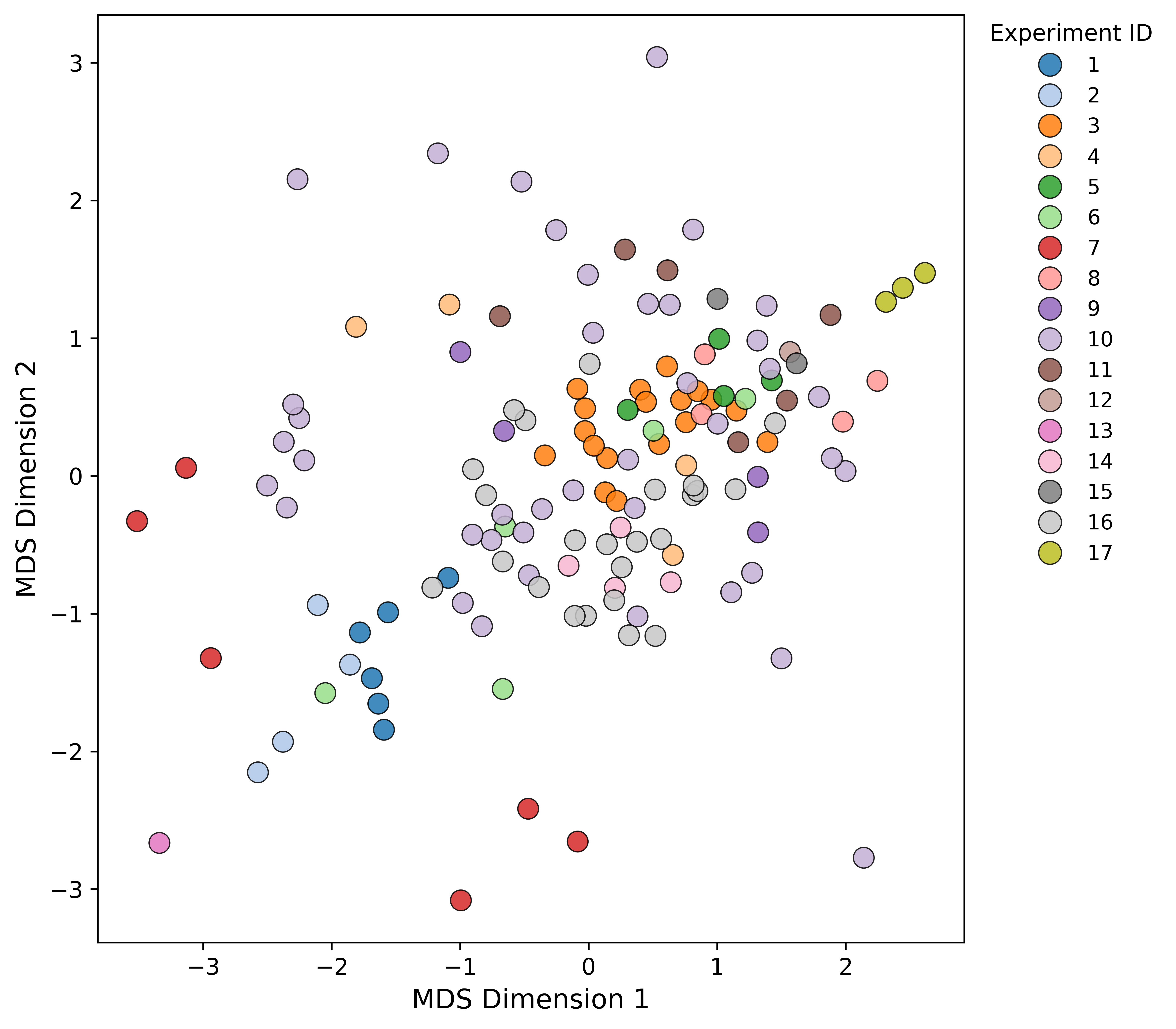}
    \caption{MDS projection of the CFD cases using the performance (latent) space.}
    \label{fig:mds_performances}
  \end{subfigure}

  \caption{Comparison of feature-based and performance-based representations of the 136 CFD cases, visualized in two dimensions using multidimensional scaling (MDS). Colors indicate the 17 distinct CFD experiments.}
  \label{fig:mds_comparison}
\end{figure}

The CFD cases investigated here can be represented by both their feature descriptions and by their performance vectors, which capture the observed case-model interactions. Figure \ref{fig:mds_comparison} visualises the corresponding representation of the selected CFD cases after mapping them onto two dimensions using multi-dimensional scaling (MDS) using the implementation provided in scikit-learn (version 1.7.0) \citep{Pedregosa2011}.

Features are mostly distinct for one CFD experiment forming experiment-specific clusters. In particular, categorical features are identical for most sub-cases of the same experiment as they typically have the same case descriptors, such as flow configuration or material system. Some experiments and their respective sub-cases (e.g. ExpID10) span a wider feature range than others (e.g. ExpID 5, 17), which can be attributed mostly to the range of their continuous features (see Table \ref{tab:caseList}). Categorical features do help identify similarities between experiments, e.g. all bubble columns experiments (ExpIDs 1, 3, 7, 12 and 13) are located on the left side of the plot. While in some experiments the performances of sub-cases are quite alike (e.g. ExpID17), in others there is diversity and a wide spread of performance points despite the feature similarity (e.g. ExpID 1 or 5).

This comparison illustrates that while feature similarity captures structural relationships between CFD cases, it does not always correspond directly to similarity in model performance. This observation motivates the use of hybrid recommender system approaches that combine feature-based case similarity with performance-based inference.

\section{Recommender model}
As a specialised class of information filtering methods, recommender systems are designed to predict which items, from a typically vast portfolio, are the most relevant to a particular user \citep{Ricci2022}. Such systems facilitate diverse decision-making processes, including music selection~\citep{schedl2019deep}, restaurant choice~\citep{alves2024regionalization}, and news article consumption~\citep{spivsak2025segment}. User–item relevance is typically quantified through explicit or implicit feedback signals, which are stored in a partially observed ratings matrix. Typically, only a tiny fraction of the entries of the rating matrix are observed. On the basis of this matrix, a recommender system infers latent preference structures and generates personalised predictions of which items are likely to be most suitable for each user.

Following earlier work on CFD closure model selection \citep{Huang2023, Haensch2025} CFD `cases' can be interpreted as users, while closure model combinations---such as specific model choices for interfacial forces, turbulence, and polydispersity---correspond to `items'. A case-item interaction produces a continuous `performance' score, an explicit and bounded value that quantifies the accuracy of simulation results when compared to validation data, as described in Section 2 and illustrated in Figure \ref{fig:performanceMatrix}. These performance values are stored in a performance matrix $R \in \mathbb{R}^{n \times m}$ where row $i$ (out of $n$) represents item $i$, column $j$ (out of $m$) represents case $j$ and the entry $(i,j)$ of $R$ the performance of item $i$ in case $j$ that evaluates the predictive quality of a particular closure model for a given case. In both structure and interpretation, this matrix is directly analogous to typical user-item rating matrices that form the backbone of conventional recommender systems.

However, the model selection problem differs from traditional recommender settings in three key aspects:
(i) labels are extremely expensive, as each performance corresponds to a fully processed CFD simulation,
(ii) side information is structured scientific metadata rather than behavioural signals, and
(iii) the objective is performance optimisation rather than user satisfaction.
These properties place the task in a distinct class of high-cost scientific recommendation problems.

The following sections briefly survey the typical recommender system methodologies before introducing the specific recommender system model employed in this paper. 

\subsection{Collaborative filtering}
Collaborative filtering bases the prediction of item relevance purely on the observed user-item interactions. For example, in movie recommendations, the system uses patterns in user–movie interactions (e.g., ratings or watch history) to predict what a user will like: if two users show similar preferences across a (small) set of movies, they are expected to rate other movies similarly as well, even those they have not watched yet. 

Translated into the CFD model selection context, the approach assumes that by collecting a sufficiently large number of CFD cases and raw performance data we can make reliable predictions about which model performs well for a certain known case. The performance representation of CFD cases from known case-item interactions collectively allows us to identify case similarity patterns. These patterns can be used for generating predictions: CFD cases that exhibit similar performance profiles for tested models are expected to behave similarly for untested model combinations.

A typical way of solving this kind of prediction problem is via matrix completion~\citep{ledent2021orthogonal,johnson2014logistic}, often implemented through matrix factorisation. Traditional matrix completion techniques, such as \textit{softImpute}~\citep{mazumder2010spectral}, often assume that the performance matrix $R$ is approximately low-rank and can be written as
\begin{equation}
R \approx UV^\top,
\end{equation}
where $U \in \mathbb{R}^{n \times d}$ and $V \in \mathbb{R}^{m \times d}$ are latent feature matrices for the $m$ closure models and $n$ CFD cases, respectively.

From this family of models, a relevant one is \emph{Gaussian Copula Imputation} (\textit{gcimpute})~\citep{Zhao2024}, which has also been used previously by the authors in CFD-style performance-matrix settings~\citep{Haensch2025} to achieve high retrieval metrics. \textit{gcimpute} decouples the modelling of marginal distributions from the modelling of dependencies by assuming that each observed value $R_{ij}$ is a monotonic transformation of a latent Gaussian variable $Z_{ij}$:
\begin{equation}
R_{ij} = F^{-1}_j \left( \Phi(Z_{ij}) \right),
\end{equation}
where $\Phi$ is the standard normal Cumulative Distribution Function (CDF) and $F^{-1}_j$ is the inverse CDF of the marginal distribution of column $j$. The latent matrix is then modelled in a low-rank form,
\begin{equation}
Z = GH^\top,
\end{equation}
which retains the matrix factorisation structure while better accommodating non-Gaussian or bounded performance measures through the copula transformation.

Although pure matrix completion can achieve significant results for case–item CFD predictions when at least a few CFD cases are known (i.e., when performance entries exist for at least some items), it cannot produce predictions for new, unseen cases. In that cold-start setting, the latent case representation cannot be determined by construction, since there are no observed entries from which to infer it. 

A pure collaborative filtering solution in such a cold-start scenario is to recommend the most popular item across all observed case–item interactions. However, this approach relies solely on popularity and does not incorporate any personalisation, resulting in identical recommendations for every new CFD case regardless of its specific features. Alternatively, a few interactions could first be computed for a new case to infer its latent representation. In practical CFD workflows, however, such exploration is often infeasible, since industrial-scale simulations can be computationally expensive. It is therefore desirable to obtain a reliable model recommendation for previously unseen cases without requiring prior simulations.

\subsection{Content-based filtering}

Content-based filtering incorporates item and user metadata to generate predictions. For instance, movie recommendation systems may use genre, cast, director, and plot keywords: if a user has liked several science-fiction films featuring a specific actor or theme, the system recommends other films with similar attributes, regardless of their rating count or release date.

In our setting, this side information is provided by the case metadata introduced in Section~\ref{sec:data}. Conditioning on these features enables model recommendations even when no performance entries are available for a new CFD case. As a result, content-based approaches naturally address the cold-start limitation of pure matrix-completion methods and allow customised model selection for previously unseen cases.

This capability is expected to become increasingly important as the dataset grows in size and diversity. A broader collection of CFD applications will span multiple material systems, geometries, boundary conditions, and flow regimes, and no single closure model is likely to perform well across all of them. Leveraging case features provides a principled way to tailor recommendations to the characteristics of each problem instance rather than relying solely on similarities in past performance matrices.

\subsection{The proposed recommender system methodology}

\begin{figure}
  \centering
  \includegraphics[width=1.0\textwidth]{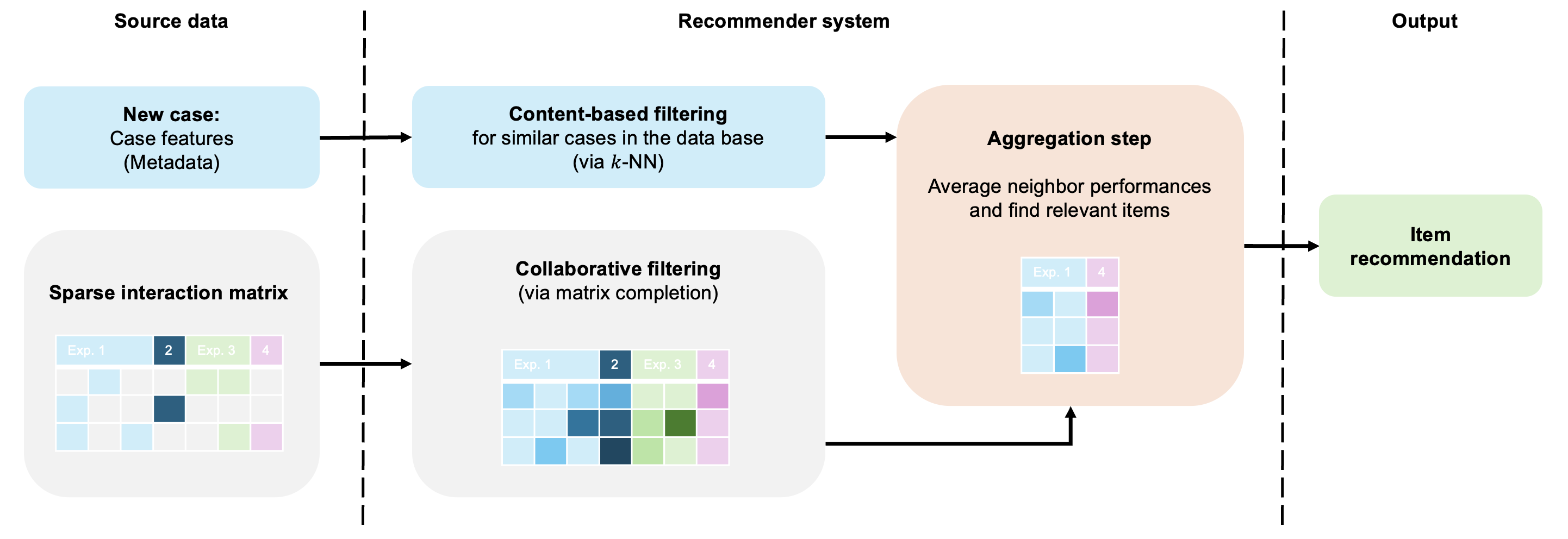}
  \caption{Schematic of the recommender system pipeline for CFD model selection for new CFD cases.}
  \label{fig:recommenderPipeline}
\end{figure}

We propose a \emph{hybrid} recommender system that combines content-based and collaborative filtering to recommend suitable CFD closure model combinations. The hybrid design is motivated by two complementary requirements: (i) enabling recommendations for \emph{new, unseen} CFD cases (cold start), and (ii) exploiting the information contained in the \emph{sparse} case--item performance matrix derived from reported simulation results. An overview of the complete pipeline is shown in Figure~\ref{fig:recommenderPipeline}.

The content-based component identifies cases that are similar \emph{in terms of case descriptors}. To this end, each CFD case is represented by a feature vector constructed from the case metadata described in Section~\ref{sec:data}. Continuous features are scaled to the range $[0,1]$ and then concatenated with the one-hot encoded categorical features (i.e. categorical values represented by binary indicator variables), resulting in a single feature vector for each case. Given two cases, their feature-based distance is computed using alternative metrics (Euclidean, Cosine, and Gower), which determines a neighbourhood structure in the feature space. The distance metrics were computed using the implementations provided in scikit-learn (version 1.7.0) \citep{Pedregosa2011}.

Formally, for a query case $q$ with feature vector $x_q$, we identify the set of $k$ nearest neighbours ($k$-NN)

\begin{equation}
\mathcal{N}_k(q) = \arg\min_{\substack{\mathcal{S}\subset \mathcal{C}\\|\mathcal{S}|=k}} \sum_{i\in \mathcal{S}} d(x_q, x_i),
\end{equation}

\noindent where $\mathcal{C}$ denotes the set of available (historical) CFD cases and $d(\cdot,\cdot)$ is the chosen distance metric. We explore different values of $k$ and distance metrics in cross-validation (Section~4.3).

In parallel, the collaborative component operates on the sparse case-item performance matrix $R_s$, whose entries correspond to observed performance outcomes for specific case-model combinations. We apply \textit{gcimpute} \citep{Zhao2024} to infer missing entries and obtain a completed (or partially completed) estimate $\hat{R}$. This imputation method was found to be the most effective for a similar data set in a previous study \citep{Haensch2025}.

The two components interact through a final aggregation step that combines (i) \emph{feature-based neighbourhood selection} and (ii) \emph{performance inference} from the completed matrix. For a query case $q$, the $k$-NN model first identifies $\mathcal{N}_k(q)$. We then retrieve, for every closure-model combination $j$, the corresponding observed or \textit{gcimpute}-predicted performance values from the neighbour cases and aggregate them into an item relevance score. Concretely, we compute
\begin{equation}
s(q,j) = \frac{1}{|\mathcal{N}_k(q)|}\sum_{i\in \mathcal{N}_k(q)} \hat{R}_{ij},
\end{equation}
where $\hat{R}_{ij}$ denotes either an observed entry (if available) or the \textit{gcimpute} prediction for neighbour case $i$ and model combination $j$. For simplicity and robustness in the small-data regime, equal weighting of neighbours was adopted; distance-weighted aggregation was tested preliminarily but did not yield systematic improvements. Ranking items by $s(q,j)$ yields the final recommendation list for the query case.

This hybrid construction preserves the strengths of both paradigms: $k$-NN over case features enables recommendations in cold-start settings (new cases with no performance history), while matrix completion increases coverage and robustness by densifying the sparse performance matrix before aggregation. Conceptually, the pipeline mirrors the reasoning of an experienced CFD practitioner: for a new case configuration, one recalls similar prior cases and then considers how different closure models performed in those contexts. The key advantage is that the proposed RS can exploit a much larger body of accumulated evidence (here comprising 13{,}600 simulations and tens of thousands of validation plots) than could reasonably be retained and processed by an individual, enabling systematic, data-driven model selection at scale.

The proposed hybrid design is motivated by the specific characteristics typical of scientific domains characterised by structured metadata, limited data availability, and high evaluation cost. First, the number of available simulation cases is relatively small (136 cases) compared to typical large-scale recommender system datasets. In such small-data regimes, highly parameterised models such as deep neural networks risk overfitting and offer limited interpretability \citep{Goodfellow2016, Zhang2019}. Distance-based similarity in the original feature space provides a robust and transparent mechanism for identifying physically related cases without introducing additional model complexity. Second, interpretability is a key requirement in scientific and engineering workflows. Practitioners need to understand why a particular model is recommended in order to build trust in the system and to diagnose potential failures. The $k$-NN component enables direct traceability of recommendations to previously studied cases with similar physical and geometric characteristics, offering an intuitive explanation mechanism that aligns with expert reasoning. Third, the performance matrix derived from CFD simulations is inherently sparse and expensive to populate. Matrix completion addresses this limitation by exploiting latent dependency structures across cases and closure model combinations, enabling reliable performance estimation even when direct observations are missing. This collaborative inference component improves robustness and coverage compared to relying solely on observed neighbour performances.

As with all data-driven decision-support systems, the proposed recommender should not be interpreted as a replacement for expert judgement. Instead, it is best viewed as an evidence-based assistance tool that highlights promising model candidates based on accumulated simulation experience, thus reducing the search space and supporting more efficient and informed expert decision-making.
\section{Evaluation methodology}
We evaluate the proposed recommender model by comparing it to various conventional baseline approaches. A fair evaluation and comparison requires careful definition of item relevance, metrics, and a proper cross-validation procedure, as described in the following sections. Table \ref{tab:notation} summarises the key notations used throughout the cross-validation procedure.

\begin{table}
\footnotesize
\caption{Notation used in the nested cross-validation and evaluation procedure.}
\label{tab:notation}
\begin{tabular}{ll}
\toprule
\textbf{Symbol} & \textbf{Description} \\
\midrule
$R$ & Fully observed ground-truth performance matrix \\
$R_s$ & Sparsified version of $R$ at sparsity level $s$ \\
$s$ & Sparsity level ($0.25,\;0.50,\;0.75,\;0.90$) \\
$\mathcal{E}$ & Set of all CFD experiments ($|\mathcal{E}|=17$) \\
$e$ & Index of the held-out test experiment \\
$v$ & Index of a validation experiment \\
$S_s$ & Validation matrix (complete - test experiment) \\
$T_s$ & Training matrix (complete - test - validation) \\
$\tilde{T}_s$ & Imputed training matrix after matrix completion \\
$\tilde{S}_s$ & Imputed validation matrix after matrix completion \\
$k$ & Number of nearest neighbors in $k$-NN \\
$RR@k$ & Reciprocal rank at cutoff $k$ (case level) \\
$RR@k^{(\mathrm{exp})}$ & Reciprocal rank averaged over cases of one experiment \\
$MRR@k_{\mathrm{val}}$ & Mean reciprocal rank on validation experiments \\
$MRR@k_{\mathrm{test}}$ & Mean reciprocal rank on test experiments \\
\bottomrule
\end{tabular}
\end{table}

\subsection{Definition of item relevance}

The definition of item relevance guides the objective of the recommender model and informs the choice of evaluation metrics \citep{Falk2019}. For a given case we define relevant items as items whose performance lies within a prescribed threshold of the best performing item. This definition of a performance threshold accounts for uncertainties in the performance computation, and deliberately includes items that are not necessarily the best, but are solid and therefore relevant alternatives. Based on prior analyses of simulation results, a performance threshold of 0.05 was adopted, for which performance differences became visually noticeable. This threshold ensures that most cases keep multiple relevant items, therefore supporting a fine-grained evaluation of results rather than forcing to converge to only a single top-performing item. In simple words, if the ground-truth best item for a case has a performance measurement of $0.90$, we consider as relevant all items whose performance is at least $0.85$.

\begin{figure}
  \centering
  \includegraphics[width=1.0\textwidth]{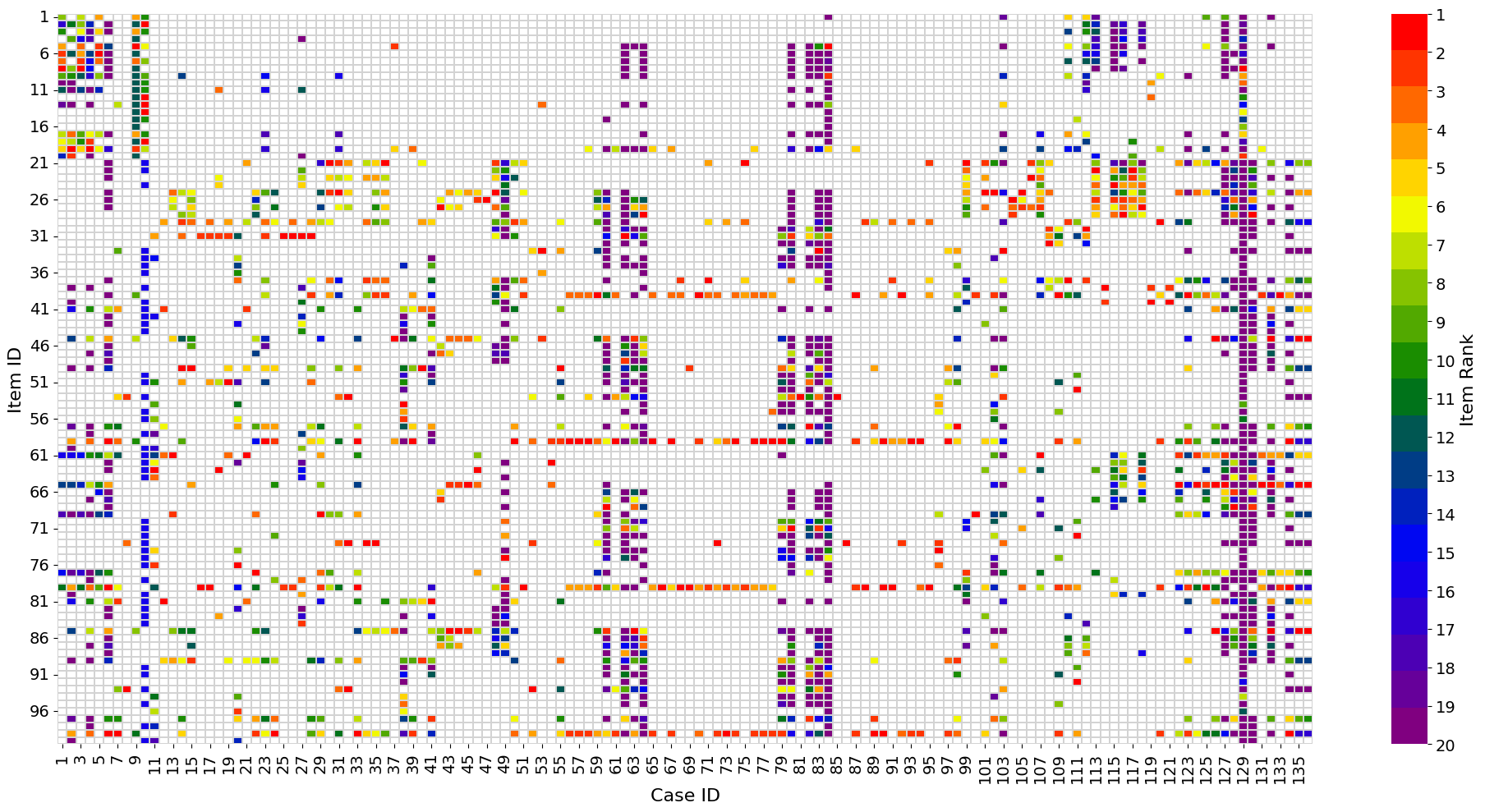}
  \caption{Relevant items achieving a performance within a 0.05 threshold of the per-case top item performance, with their ranking indicated by colour coding.}
  \label{fig:rankedMatrix}
\end{figure}

Figure \ref{fig:rankedMatrix} illustrates the relevant items for each case extracted from the ground-truth matrix. 

\subsection{Evaluation metrics}

Based on the relevant items, we make use of the ranking metric reciprocal rank (RR) at the top \textit{k} items, which is generally computed as:

\begin{equation}
    \label{eqn: RR}
    RR@k = 
    \begin{cases}
\frac{1}{rank_i}       & \text{if relevant item $i$ is among the top \textit{k}} \\
0                      & \text{if no relevant item is in the top \textit{k}}.
\end{cases}
\end{equation}

This metric rewards recommenders that rank relevant items higher, with better scores when relevant items appear closer to the top of the list. 
We then compute the reciprocal rank (RR) at the top \textit{k} items for each individual experiment, i.e. as the average RR@k over all sub-cases belonging to the same experiment:

\begin{equation}
    \label{eqn: RR_exp}
    RR@k^{(exp)} = \frac{1}{N_{cases}}\sum_{c=1}^{N_{cases}} RR@k^{(c)},
\end{equation}

where $N_{cases}$ denotes the number of CFD sub-cases in the experiment, and $RR@k^{(c)}$ is the reciprocal rank of the relevant item within the top \textit{k} predictions for case c. Thus, we guarantee a fair representation of each individual CFD experiment during the evaluation process regardless of the number of its sub-cases.

For evaluating our model selection recommender, we define the mean reciprocal rank (MRR) at the top \textit{k} items as the average over all experiments:

\begin{equation}
    \label{eqn: MRR}
    MRR@k = \frac{1}{N_{exp}}\sum_{e=1}^{N_{exp}} RR@k^{(exp)}.
\end{equation}

Note that the experiments here include all train or validation experiments depending on the step in the cross-validation described below. 

\subsection{Testing and Cross-validation}
The testing and cross-validation procedure adopted in this work is designed to assess the generalisation capability of the recommender model, that is the model's ability to extend its predictive accuracy to completely unobserved data. Although the ground-truth matrix in our data set is fully observed, it is deliberately sparsified to emulate realistic scenarios in which only a limited number of performance evaluations are available. This allows us to systematically investigate the potential of exploiting sparse matrix data.

As pointed out in Section~2 and illustrated in Figure~\ref{fig:mds_comparison}, cases belonging to the same CFD experiment exhibit highly similar feature characteristics and, more importantly, similar performance behaviour. In practice, however, a new CFD experiment is typically simulated without prior knowledge of its sub-cases and not a single performance value exists for the entire experiment. Consequently, a rigorous evaluation procedure must ensure that all cases associated with a given test experiment remain entirely unseen during both model training and hyperparameter tuning. This approach ensures a robust and reliable assessment of model performance in practical applications, where accurate extrapolation beyond previously observed CFD experiments is essential. 

To this end, we employ a nested cross-validation strategy for model evaluation, as outlined below in Algorithm~\ref{alg:nested_cv}. Unlike typical evaluation protocols in recommender systems that randomly hold out interactions, we simulate a true scientific cold-start scenario by withholding entire experiments, ensuring that no performance information from the test scenario is available during training. To the best of our knowledge, this is the first study to demonstrate CFD model recommendation under strict experiment-level cold-start conditions. In brief, the fully observed ground-truth matrix is systematically sparsified to emulate varying data availability, after which each of the 17 CFD experiments is treated as a held-out test set. Model hyperparameters are selected via an inner cross-validation loop on the remaining 16 experiments, while matrix completion and $k$-NN-based prediction are performed exclusively on data not containing the test experiment. The nested structure of the evaluation procedure, including the separation between test evaluation and inner hyperparameter tuning, is illustrated in Figure~\ref{fig:cv_flow}.

A more detailed procedural description of this nested cross-validation is provided in the Appendix.

\begin{algorithm}
\caption{Nested cross-validation for recommender model evaluation.}
\label{alg:nested_cv}
\footnotesize
\begin{algorithmic}[1]
\REQUIRE Fully observed ground-truth matrix $R$, sparsity levels $s$, test experiments $\mathcal{E}$
\FOR{each sparsity level $s \in \{0.25, 0.50, 0.75, 0.90\}$}
    \FOR{each random realization $R_s$ of sparsified $R$}
        \FOR{each test experiment $e \in \mathcal{E}$}
            \STATE Remove experiment $e$ from $R_s$ to form validation matrix $S_s$
            \FOR{each validation experiment $v \neq e$}
                \STATE Remove $v$ from $S_s$ to form training matrix $T_s$
                \STATE Impute missing entries in $T_s$ using \textit{gcimpute} to obtain $\tilde{T}_s$
                \FOR{each $k$-NN configuration}
                    \STATE Predict performances for validation sub-cases using k nearest neighbors
                    \STATE Compute the average $RR@3^{(v)}$
                \ENDFOR
            \ENDFOR
            \STATE Select best $k$-NN configuration based on $MRR@3_{\mathrm{val}}$
            \STATE Impute missing entries in $S_s$ using \textit{gcimpute} to obtain $\tilde{S}_s$
            \STATE Predict performance for test sub-cases of experiment $e$ 
            \STATE Compute $RR@3^{(e)}$
        \ENDFOR
    \ENDFOR
\ENDFOR
\STATE Compute final $MRR@3_{\mathrm{test}}$ by averaging over all test experiments
\end{algorithmic}
\end{algorithm}

\newpage

\begin{figure}
\centering
\footnotesize
\begin{tikzpicture}[
    node distance=1.6cm,
    every node/.style={draw, rounded corners, align=center},
    arrow/.style={->, thick}
]

\node (start) {Fully observed matrix $R$};

\node (sparse) [below of=start] {Generate sparse matrices $R_s$\\ ($s=0.25,0.50,0.75,0.90$)};

\node (test) [below of=sparse] {Select test experiment $e$\\ (all sub-cases removed)};

\node (branch) [below of=test] {Validation matrix $S_s$\\ (complete $\setminus$ test)};

\node (imputeS) [below of=branch, yshift=-3.6cm] {Matrix completion\\ $\tilde{S}_s$};

\node (testpred) [below of=imputeS] {$k$-NN prediction\\ (test experiment)};

\node (eval) [below of=testpred] {$RR@3^{(e)}$};

\node (final) [below of=eval] {Average over experiments\\ $MRR@3_{\mathrm{test}}$};

\node (train) [right=4.5cm of branch] {Training matrix $T_s$\\ (complete $\setminus$ test $\setminus$ validation)};

\node (imputeT) [below of=train] {Matrix completion\\ $\tilde{T}_s$};

\node (knn) [below of=imputeT] {$k$-NN prediction\\ (validation cases)};

\node (valeval) [below of=knn] {$RR@3^{(v)}$};

\node (select) [below of=valeval] {Select best $k$-NN\\ via $MRR@3_{\mathrm{val}}$};

\draw[arrow] (start) -- (sparse);
\draw[arrow] (sparse) -- (test);
\draw[arrow] (test) -- (branch);

\draw[arrow] (branch) -- (train);
\draw[arrow] (train) -- (imputeT);
\draw[arrow] (imputeT) -- (knn);
\draw[arrow] (knn) -- (valeval);
\draw[arrow] (valeval) -- (select);

\draw[arrow] (select.west) -- ++(-2.6cm,0) |- (imputeS.east);

\draw[arrow] (imputeS) -- (testpred);
\draw[arrow] (testpred) -- (eval);
\draw[arrow] (eval) -- (final);

\node[draw, dashed, fit=(train)(select), inner sep=8pt, label=above:{\textbf{Inner cross-validation loop}}] {};

\end{tikzpicture}
\caption{Flow diagram of the nested cross-validation procedure. The inner cross-validation loop (right) is used for $k$-NN hyperparameter selection and feeds back into the main test loop after selecting the optimal configuration.}
\label{fig:cv_flow}
\end{figure}
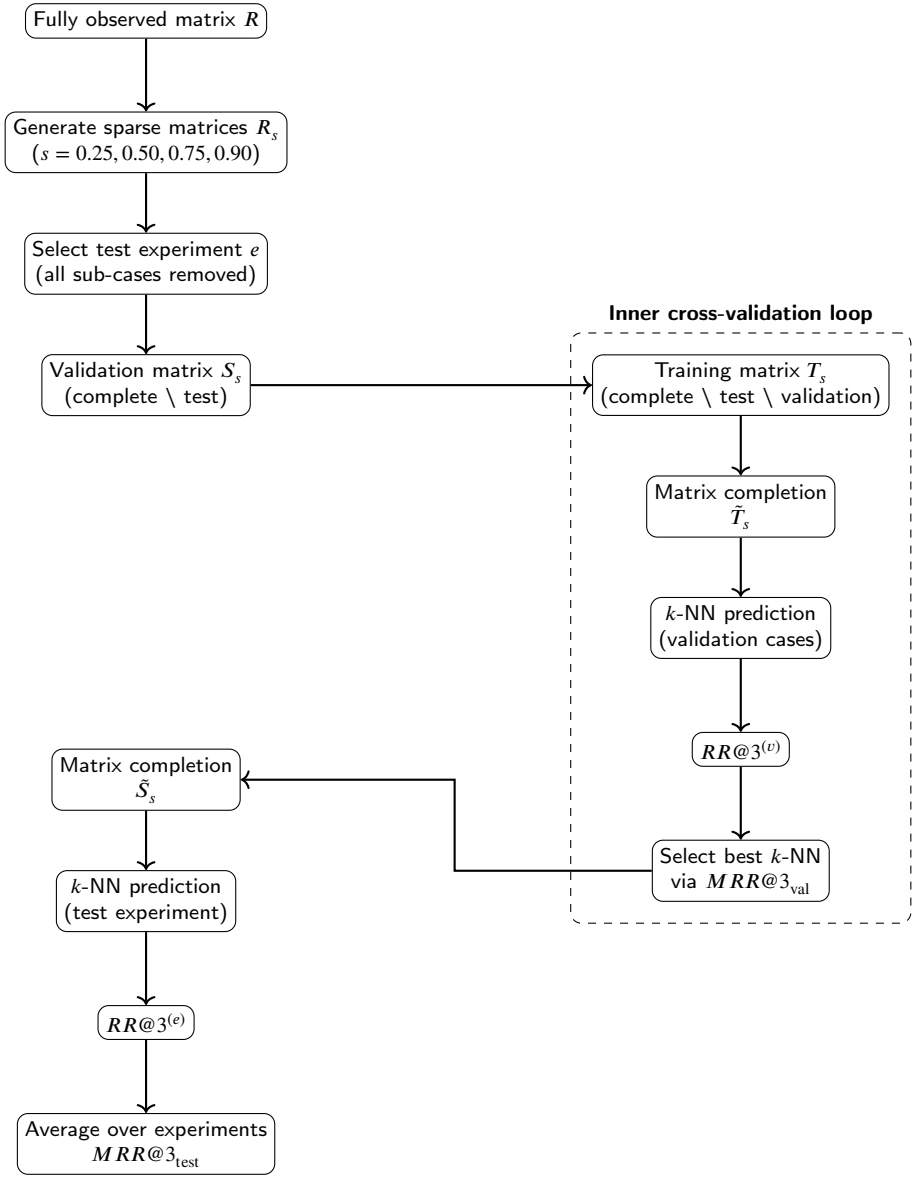
\section{Results}

\subsection{Optimised $k$-NN configurations}

Distinct optimised $k$-NN configurations are obtained for each sparsity level and for each individual CFD test experiment. Table~\ref{tab:best_knn} summarises the optimal $k$-NN hyperparameter settings for the 17 test experiments at a sparsity level of $s = 0.75$. For each test experiment, comprising a different number of cases, the optimal distance metric and the number of neighbours, $k$, were identified by $k$-fold cross-validation using the remaining 16 experiments as validation data.
Although cosine distance with $k = 10$ is selected most frequently at $s = 0.75$, no consistent trend with respect to sparsity is observed in either the average optimal number of neighbours (computed across all test experiments) or the preferred distance metric.

\begin{table}
\centering
\caption{List of the best $k$-NN parameters for the individual test experiments for $s=0.75$, reporting the test-set mean $RR@k$ values averaged over 100 sparse matrix versions. A more detailed version can be found as Table \ref{tab:best_knn_ci} in the Appendix.}
\label{tab:best_knn}
\begin{tabular}{ccccc}
\toprule
\textbf{Exp.ID} & \textbf{Metric} & \textbf{k} & \textbf{RR@1} & \textbf{RR@3} \\
\midrule
1  & cosine    & 10 & 0.325 & 0.437 \\
2  & cosine    & 10 & 0.235 & 0.312 \\
3  & euclidean & 10 & 0.189 & 0.273 \\
4  & cosine    & 10 & 0.677 & 0.813 \\
5  & cosine    & 10 & 0.560 & 0.692 \\
6  & cosine    & 10 & 0.496 & 0.641 \\
7  & cosine    & 10 & 0.292 & 0.416 \\
8  & cosine    & 10 & 0.725 & 0.835 \\
9  & euclidean & 50 & 0.072 & 0.138 \\
10 & cosine    & 10 & 0.476 & 0.594 \\
11 & gower     & 50 & 0.273 & 0.358 \\
12 & cosine    & 10 & 0.520 & 0.683 \\
13 & cosine    & 10 & 0.850 & 0.922 \\
14 & cosine    & 10 & 0.287 & 0.399 \\
15 & cosine    & 10 & 0.545 & 0.652 \\
16 & cosine    & 10 & 0.485 & 0.568 \\
17 & cosine    & 10 & 0.973 & 0.987 \\
\bottomrule
\end{tabular}
\end{table}

\subsection{Comparison with data-driven and expert-designed baselines}

We now compare the hybrid recommender results against two different conventional recommendation approaches:

(i) a \textbf{data-driven popularity baseline} selecting the popular item with the highest mean performance across experiments from the observed matrix entries, which represents a typical case-independent benchmark when evaluating recommender systems, and

(ii) an \textbf{expert-designed reference model}, representing a single, generally applicable closure model combination following a ``one-model-fits-all'' paradigm commonly applied in practice. Although multiple universal model sets exist in the literature, this study is restricted to the model set highlighted in Table \ref{tab:itemPortfolio} and provided by the publicly available repository of \cite{HZDRcode}.

\subsubsection{Ranking metrics}

\begin{figure}
  \centering
  \begin{subfigure}[t]{0.95\textwidth}
    \centering
    \includegraphics[width=\textwidth]{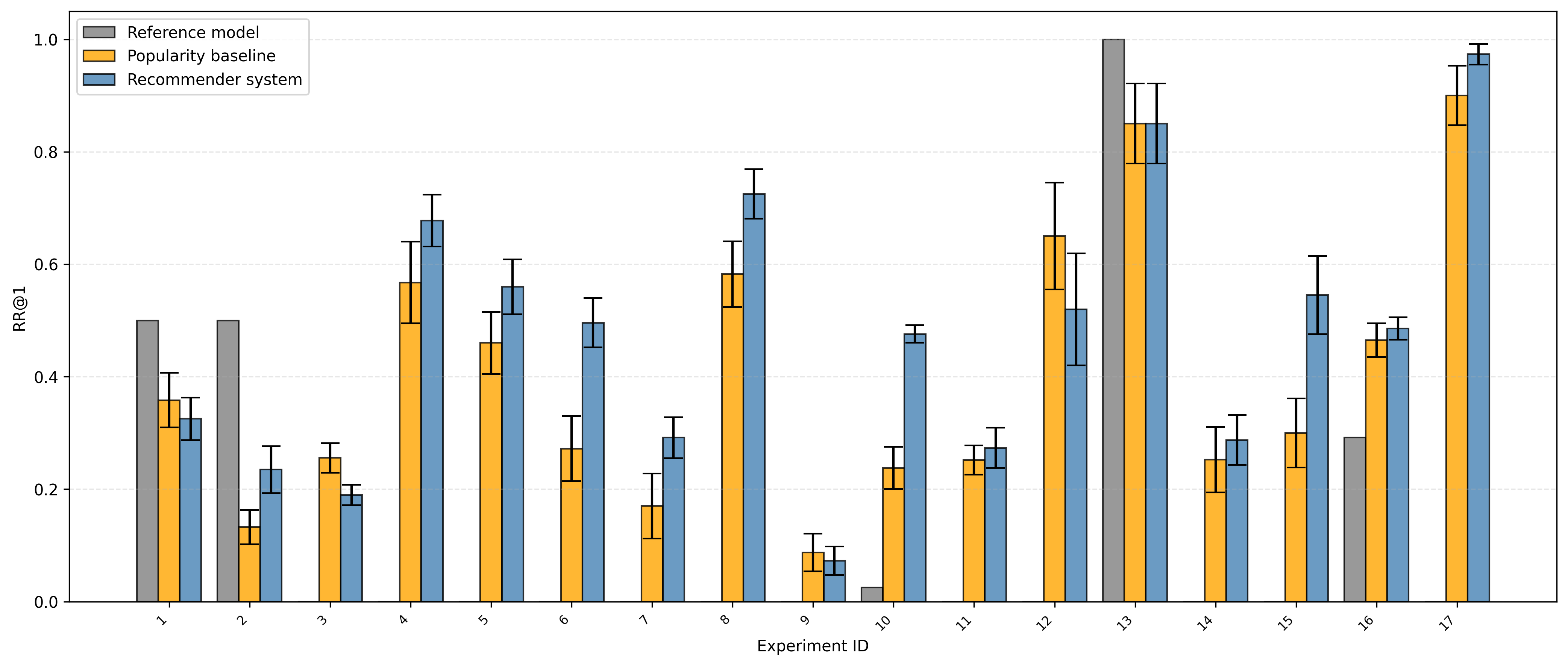}
    \caption{\textit{RR@1} per experiment for the reference model, popular item, and recommender prediction derived from a 75\% sparse matrix.}
    \label{fig:mrr1PerExp}
  \end{subfigure}

  \vspace{0.5em}

  \begin{subfigure}[t]{0.95\textwidth}
    \centering
    \includegraphics[width=\textwidth]{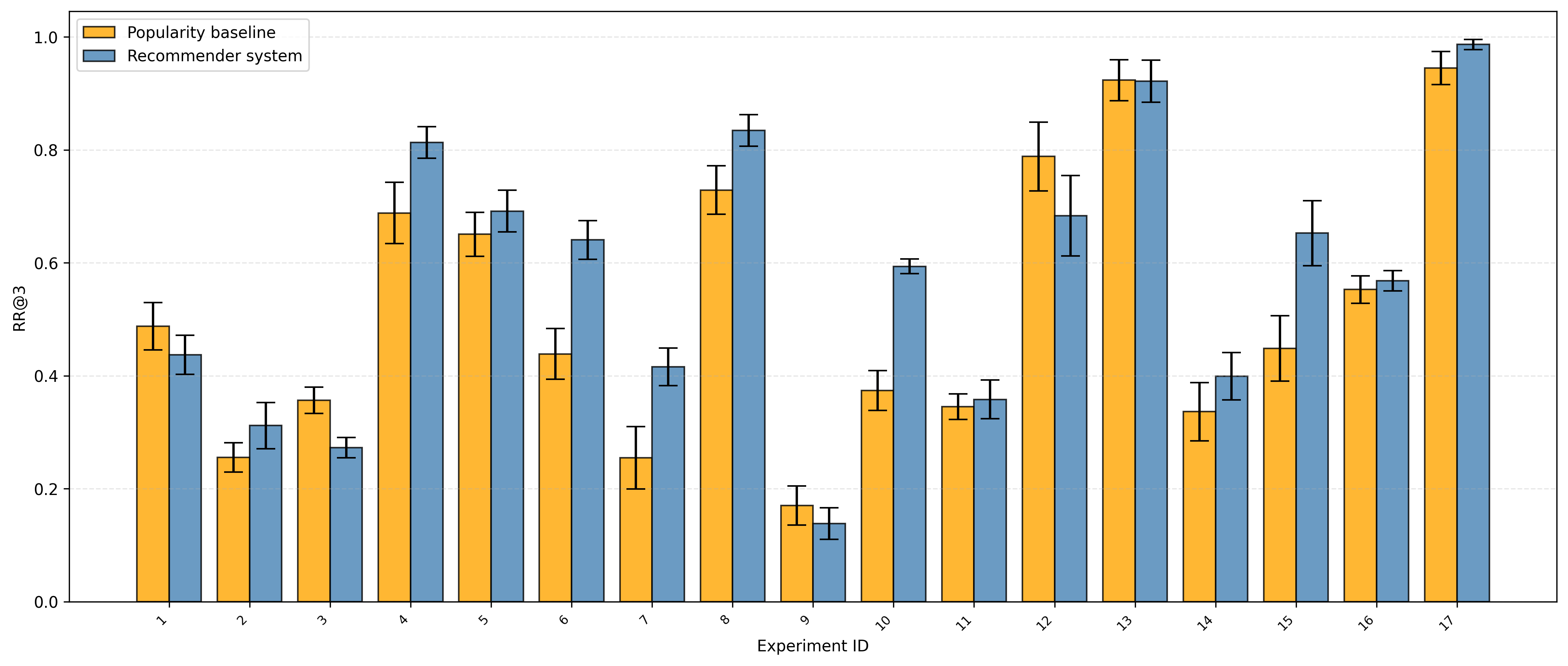}
    \caption{\textit{RR@3} per experiment for the popular item and the recommender prediction derived from a 75\% sparse matrix.}
    \label{fig:mrr3PerExp}
  \end{subfigure}

  \caption{Reciprocal rank (RR) per test experiment (averaged over sub-cases), evaluated on a 75\% sparse performance matrix with 95\% confidence intervals. Panel (a) compares \textit{RR@1} across reference model, popular-item, and recommender model, while panel (b) reports the corresponding \textit{RR@3} results for the popular-item and the recommender model.}
  \label{fig:mrrPerExpCombined}
\end{figure}

Table \ref{tab:best_knn} reports the results for \textit{RR@1} and \textit{RR@3} applying the optimised $k$-NN configurations for the respective test experiments. Figure~\ref{fig:mrrPerExpCombined} illustrates this per-experiment prediction accuracy at $s=0.75$ in terms of $\textit{RR@1}$ and $\textit{RR@3}$. The lowest performance is observed for Experiment 9, which contains only a small number of relevant closure models, making reliable recommendations intrinsically difficult. In contrast, Experiments 13 and 17, which exhibit a moderate number of relevant items, yield the highest scores for the popularity baseline and the $k$-NN-based recommendations. A detailed overview of the results across all sparsity levels for each individual test experiment is given in Table \ref{tab:best_knn_ci} of the Appendix.

Compared to the popularity baseline, the hybrid recommender achieves higher $\textit{RR@k}$ values in 12 out of the 17 test experiments for s=0.75. The reference model outperforms the other approaches in Experiments 1, 2, and 13, but otherwise frequently fails to enter the relevant-item set, highlighting the limitations of universal closure assumptions. Note that we cannot include the reference model result for the \textit{RR@3} as the metric computation requires three predicted items, while the reference model comprises of one item only.

Table \ref{tab:mrr_sparsity_combined} reports the mean reciprocal rank $\textit{MRR@1}$ and $\textit{MRR@3}$ across all test experiments as a function of sparsity and compares the proposed hybrid recommender system (RS) against several baselines. In addition to the popularity baseline computed from observed entries (Pop), we include a second popularity variant derived from the matrix-completed performance matrix (MC). This allows us to disentangle the effects of collaborative imputation from those of feature-based neighbourhood modelling. Results are further contrasted with the expert-designed reference model and with random recommendations, which yield much lower sparsity-independent performance levels.

As expected, the popularity baseline degrades monotonically with increasing sparsity and trends toward random performance in the limit of very sparse observations. The popularity variant based on matrix completion (MC) partially mitigates this degradation, indicating that collaborative inference via matrix completion recovers useful global performance structure even when direct observations are scarce. However, MC selects a single globally best item independent of case-specific characteristics.

\begin{figure}
  \centering
  \includegraphics[width=0.5\textwidth]{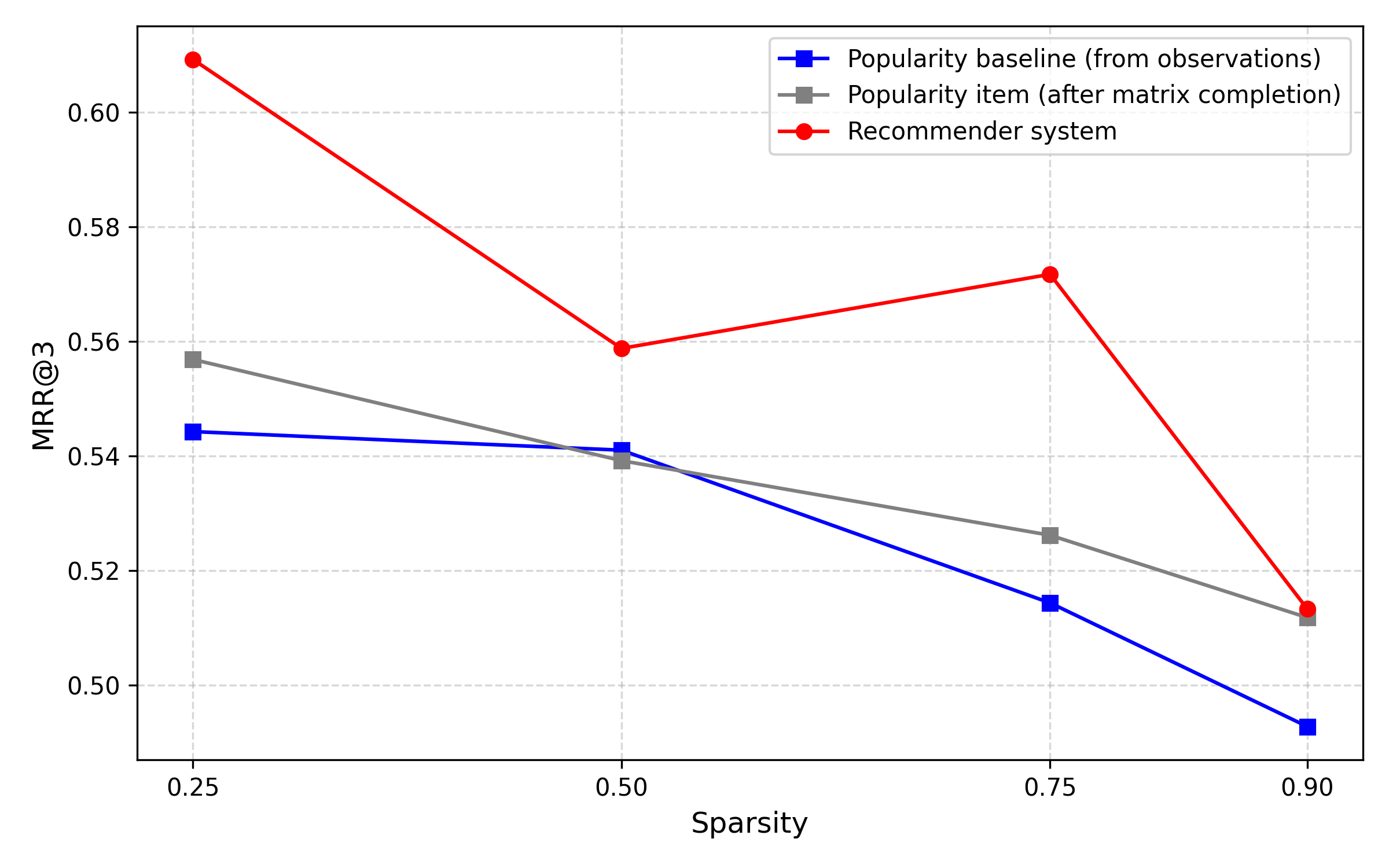}
  \caption{MRR@3 for the popularity baseline from observed entries and after matrix completion compared to the recommender system predictions, plotted over the different sparsity levels explored.}
  \label{fig:mrr3_over_sparsity}
\end{figure}

The proposed hybrid recommender consistently achieves the highest performance across all sparsity levels. Notably, the wide performance gap between the recommender and both popularity variants from $s=0.25$ to $s=0.75$ demonstrates that the integration of feature-based $k$-NN neighbourhoods provides substantial additional predictive power beyond matrix completion alone. The ablation results shown in Figure \ref{fig:mrr3_over_sparsity} make this effect explicit and isolate the contribution of metadata-driven neighbourhood modelling beyond collaborative imputation. While matrix completion improves global ranking quality, the dominant performance gains arise from neighbour selection, which enables case-specific adaptation rather than global popularity selection.

At the highest sparsity level ($s=0.90$), the margin between RS and MC narrows, which might reflect the intrinsic difficulty of reliable ranking under extremely limited observations. Nevertheless, even in this regime, the hybrid recommender remains competitive and avoids the sharp degradation observed for the purely observational popularity baseline.

Overall, the ablation analysis confirms that the performance improvements of the proposed approach cannot be attributed solely to imputation of missing entries. Instead, they stem from the principled combination of collaborative inference and content-based case similarity.

\begin{table}
\centering
\caption{MRR@1 and MRR@3 of the popularity baseline from observed entries (Pop) and from imputed entries after matrix completion (MC), compared against predictions of the proposed hybrid recommender system (RS) across different sparsity levels. Results are also shown for the expert-designed reference model, and random recommendations.}
\label{tab:mrr_sparsity_combined}
\begin{tabular}{c|ccc|ccc}
\toprule
\textbf{Sparsity} 
& \multicolumn{3}{c|}{\textbf{MRR@1}} 
& \multicolumn{3}{c}{\textbf{MRR@3}} \\
\cmidrule(lr){2-4} \cmidrule(lr){5-7}
& \textbf{Pop} & \textbf{MC} & \textbf{RS}
& \textbf{Pop} & \textbf{MC} & \textbf{RS} \\
\midrule
0.25 & 0.443 & 0.461 & 0.524 & 0.544 & 0.557 & 0.609 \\
0.50 & 0.429 & 0.422 & 0.459 & 0.541 & 0.539 & 0.559 \\
0.75 & 0.400 & 0.405 & 0.470 & 0.514 & 0.526 & 0.572 \\
0.90 & 0.374 & 0.396 & 0.400 & 0.493 & 0.512 & 0.513 \\
\midrule
Reference & -- & -- & 0.136 & -- & -- & -- \\
Random    & -- & -- & 0.165 & -- & -- & 0.254 \\
\bottomrule
\end{tabular}
\end{table}

\newpage
\subsubsection{Performance regret}
While the $\textit{MRR@k}$ metrics quantify the ranking quality, they do not capture the magnitude of performance loss when a sub-optimal model is selected. To address this we additionally evaluate \textit{Regret}, defined as the loss in performance relative to the true best item for a given case. Lower \textit{Regret} implies reduced risk of selecting physically misleading closure models, which is critical in engineering decision support where sub-optimal choices can propagate into design errors.

Let $p_c^\star$ denote the observed performance of the true best item (closure model combination) for a case $c$ in the fully observed ground-truth data, $p_c^{\text{RS}}$ the performance of the item recommended via recommender system, $p_c^{\text{Pop}}$ the performance of the popular item from observed entries, and $p_c^{\text{Ref}}$ the performance of the reference model. We define the \textit{Regret} for our recommender prediction, the popular and reference items as:

\begin{align}
\mathrm{Regret}_c^{\text{RS}} &= p_c^\star - p_c^{\text{RS}}, \\
\mathrm{Regret}_c^{\text{Pop}} &= p_c^\star - p_c^{\text{Pop}}, \\
\mathrm{Regret}_c^{\text{Ref}} &= p_c^\star - p_c^{\text{Ref}}.
\end{align}

\begin{figure}
  \centering
  \includegraphics[width=0.95\textwidth]{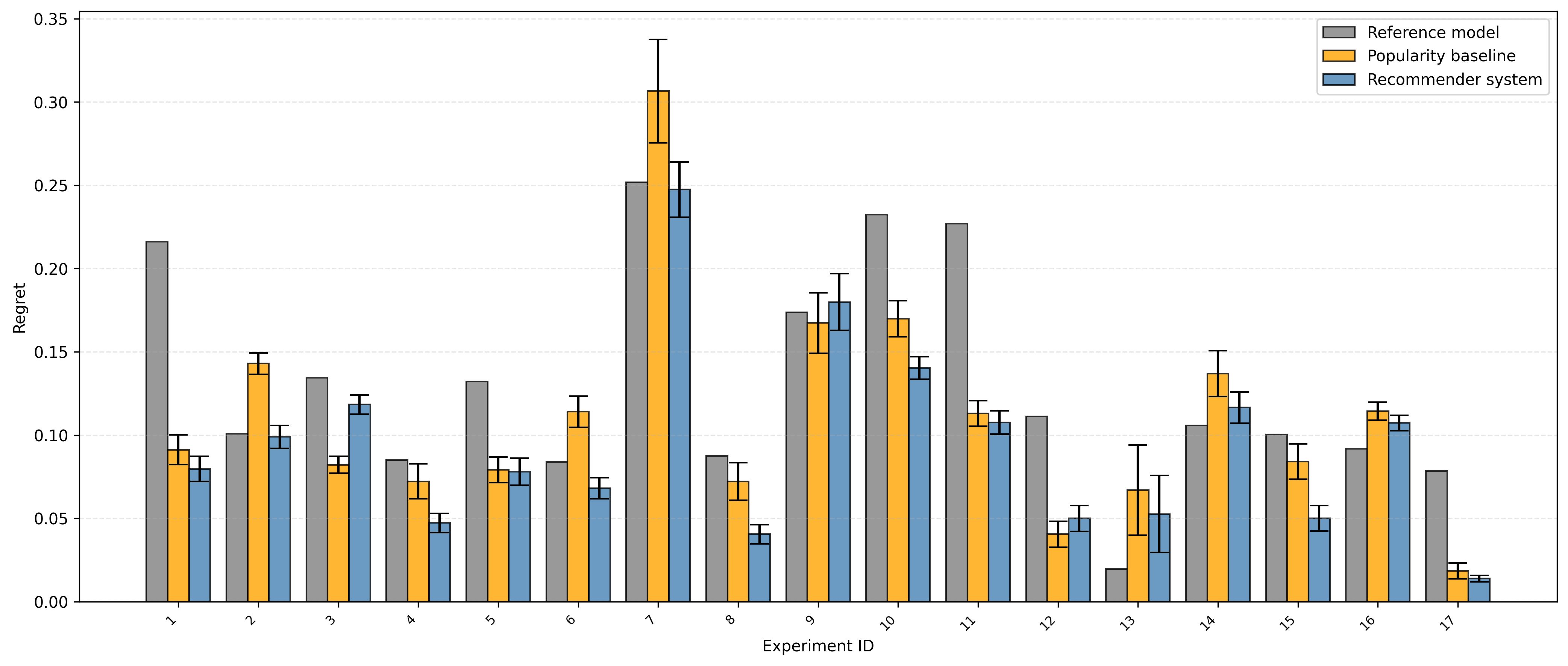}
  \caption{\textit{Regret} per experiment (consisting of a certain number of CFD sub-cases) for the reference model, the recommender-predicted and popular best item derived from a 75\% sparse matrix with 95\% confidence intervals.}
  \label{fig:regretPerExp}
\end{figure}

Averaged over sub-cases belonging to the same experiment we get the \textit{Regret} comparison illustrated for a sparsity level of $s=0.75$ in Figure \ref{fig:regretPerExp}. Compared with the popularity baseline, the recommender predictions achieve lower \textit{Regret} in most test experiments (14 out of 17), and this improvement is statistically significant. Particular improvements versus the expert-designed reference model can be reported for Experiment IDs 1, 5, 10, 11, 12 and 17. Experiment 7 exhibits high \textit{Regret} for all methods, which can be explained by the high sensitivity of this particular experiment regarding unstable staggering solutions, for which the performance was set to zero.

\begin{figure}
  \centering
  \includegraphics[width=0.5\textwidth]{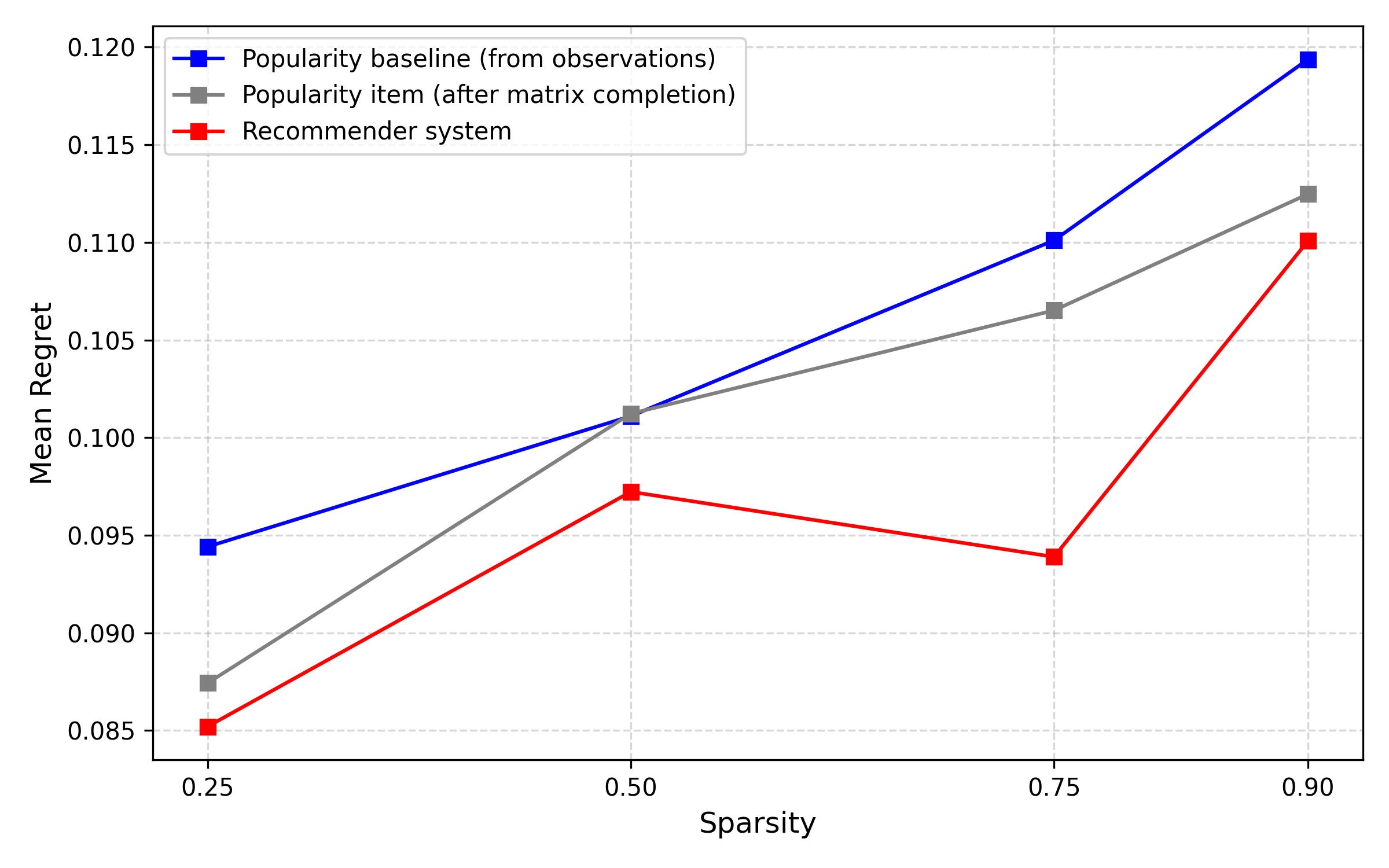}
  \caption{Average \textit{Regret} for the popularity baseline from observed entries and after matrix completion compared to the recommender system predictions, plotted over the different sparsity levels.}
  \label{fig:regret_over_sparsity}
\end{figure}

Table \ref{tab:regret_sparsity} and Figure \ref{fig:regret_over_sparsity} present the average \textit{Regret} values across all test experiments as a function of sparsity. The hybrid recommender system consistently yields lowest mean \textit{Regret} values, indicating the most reliable selection of high-performing closure models. The superiority of the RS is particularly pronounced at moderate sparsity levels ($s=0.25$ to $s=0.75$), where it consistently outperforms both the popularity baseline derived from observed entries (Pop) and the popularity variant obtained after matrix completion (MC). As shown in Figure \ref{fig:regret_over_sparsity}, the gap between RS and the case-independent baselines increases with sparsity up to $s=0.75$, indicating that metadata-driven neighbourhood modelling becomes increasingly beneficial as direct performance observations become scarce.

The popularity variant after matrix completion (MC) improves over the purely observational baseline at higher sparsity levels, confirming that collaborative imputation successfully reconstructs aspects of the latent performance structure. However, MC remains systematically inferior to RS. This demonstrates that imputation alone does not suffice for robust model selection: while matrix completion improves global performance estimates, it lacks the case-specific discrimination enabled by feature-based $k$-NN neighbourhood selection. The ablation therefore provides clear evidence that the reduction in \textit{Regret} cannot be attributed solely to matrix completion, but results from the principled integration of collaborative inference with structured metadata.

At the highest sparsity level ($s=0.90$), all methods exhibit increased \textit{Regret}, reflecting the intrinsic difficulty of reliable ranking under extreme data scarcity. In this regime, the performance gap between RS and MC narrows, suggesting that when observational information becomes severely limited, the relative advantage of neighbourhood modelling diminishes. Nevertheless, the hybrid recommender maintains the lowest overall \textit{Regret}, demonstrating stable behaviour even under highly constrained information conditions.

The expert-designed reference model, while not competitive in ranking-based metrics, achieves moderate \textit{Regret} and substantially outperforms random recommendation (see Table \ref{tab:regret_sparsity}). While the recommender model does produce the lowest \textit{Regret} values for all sparsity levels, the observed performance benefits compared to the robust reference model are moderate at the highest sparsity investigated as part of this study. This behaviour indicates that the reference model constitutes a solid, conservative and robust default choice for a given new CFD case, avoiding catastrophic performance degradation even though it lacks adaptivity. Importantly, however, the hybrid recommender consistently achieves lower \textit{Regret} than the reference model, confirming that data-driven, case-specific adaptation can reduce performance loss without sacrificing robustness.

Overall, the regret analysis strengthens the central claim of this study: the hybrid integration of collaborative matrix completion and metadata-driven neighbourhood modelling not only improves ranking quality but also reduces the practical risk of selecting substantially sub-optimal closure models. In high-cost and high-risk scientific simulation settings, where each erroneous selection may entail days of computational expense and wrong design choices, such risk reduction is arguably more critical than marginal improvements in ranking metrics alone. 

\begin{table}
\centering
\caption{Mean \textit{Regret} of the recommender predictions (RS), of the popularity baseline (Pop) and the popular item after matrix completion (MC) across different sparsity levels, compared against reference model and random recommendations.}
\label{tab:regret_sparsity}
\begin{tabular}{cccc}
\toprule
\textbf{Sparsity} & \textbf{Pop} & \textbf{MC} & \textbf{RS}\\
\midrule
0.25      & 0.094 & 0.087 & 0.085  \\
0.50      & 0.101 & 0.101 & 0.097  \\
0.75      & 0.110 & 0.107 & 0.094  \\
0.90      & 0.119 & 0.112 & 0.110  \\
\bottomrule
\textbf{Reference} & -- & -- & 0.131 \\
\bottomrule
\textbf{Random}  & -- & -- & 0.192
\end{tabular}
\end{table}

\begin{figure}
    \centering

    \begin{flushleft}
    \begin{subfigure}[b]{0.45\textwidth}
        \includegraphics[width=\textwidth]{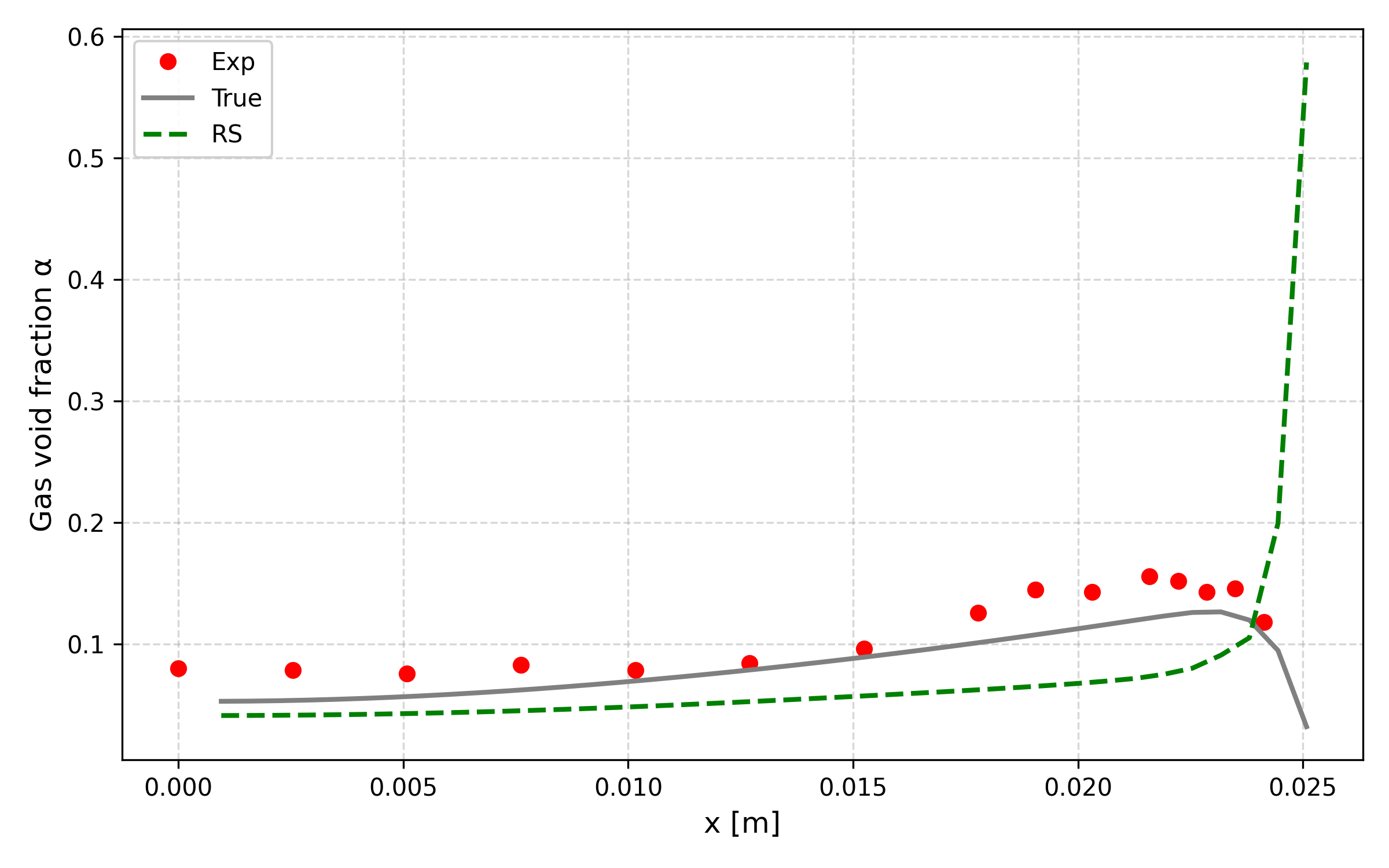}
        \caption{Gas void fraction}
        \label{fig:32_alpha}
    \end{subfigure}
    \hspace{0.5cm}
    \begin{subfigure}[b]{0.45\textwidth}
        \includegraphics[width=\textwidth]{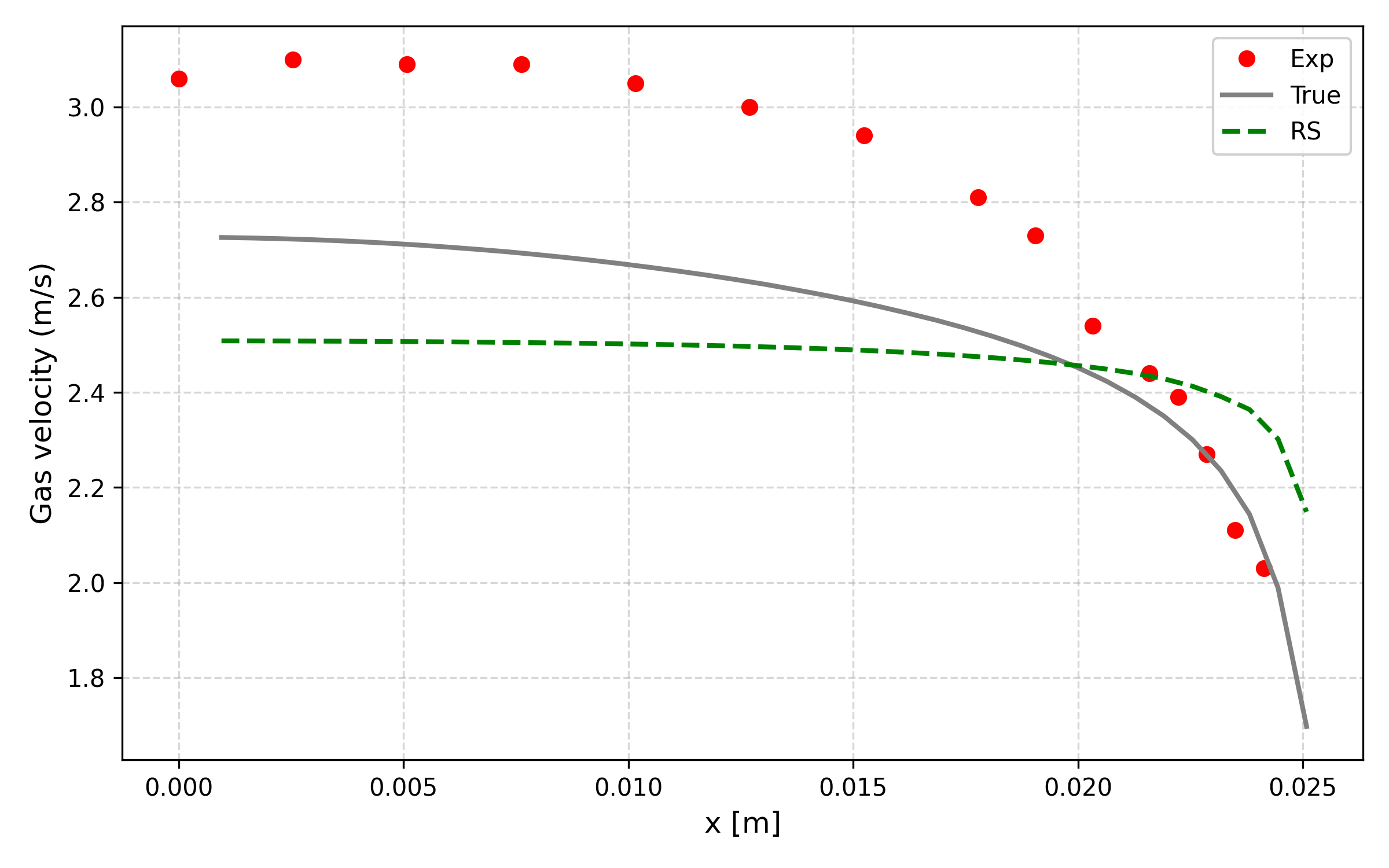}
        \caption{Gas velocity}
        \label{fig:32_gasVel}
    \end{subfigure}
    \end{flushleft}

    \vspace{0.4cm}

    \begin{flushleft}
    \begin{subfigure}[b]{0.45\textwidth}
        \includegraphics[width=\textwidth]{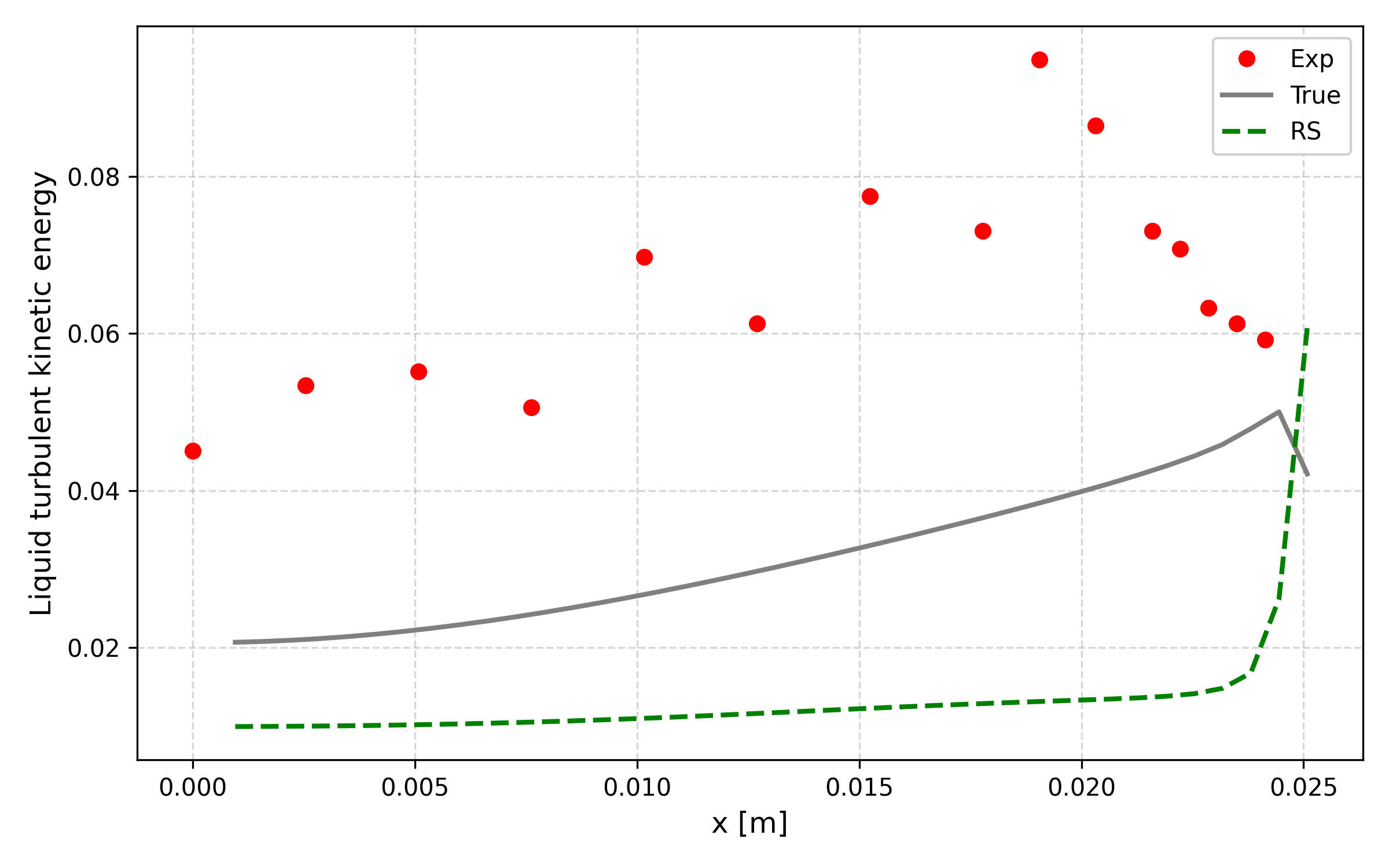}
        \caption{Liquid turb. kinetic energy}
        \label{fig:32_liqKe}
    \end{subfigure}
    \hspace{0.5cm}
    \begin{subfigure}[b]{0.45\textwidth}
        \includegraphics[width=\textwidth]{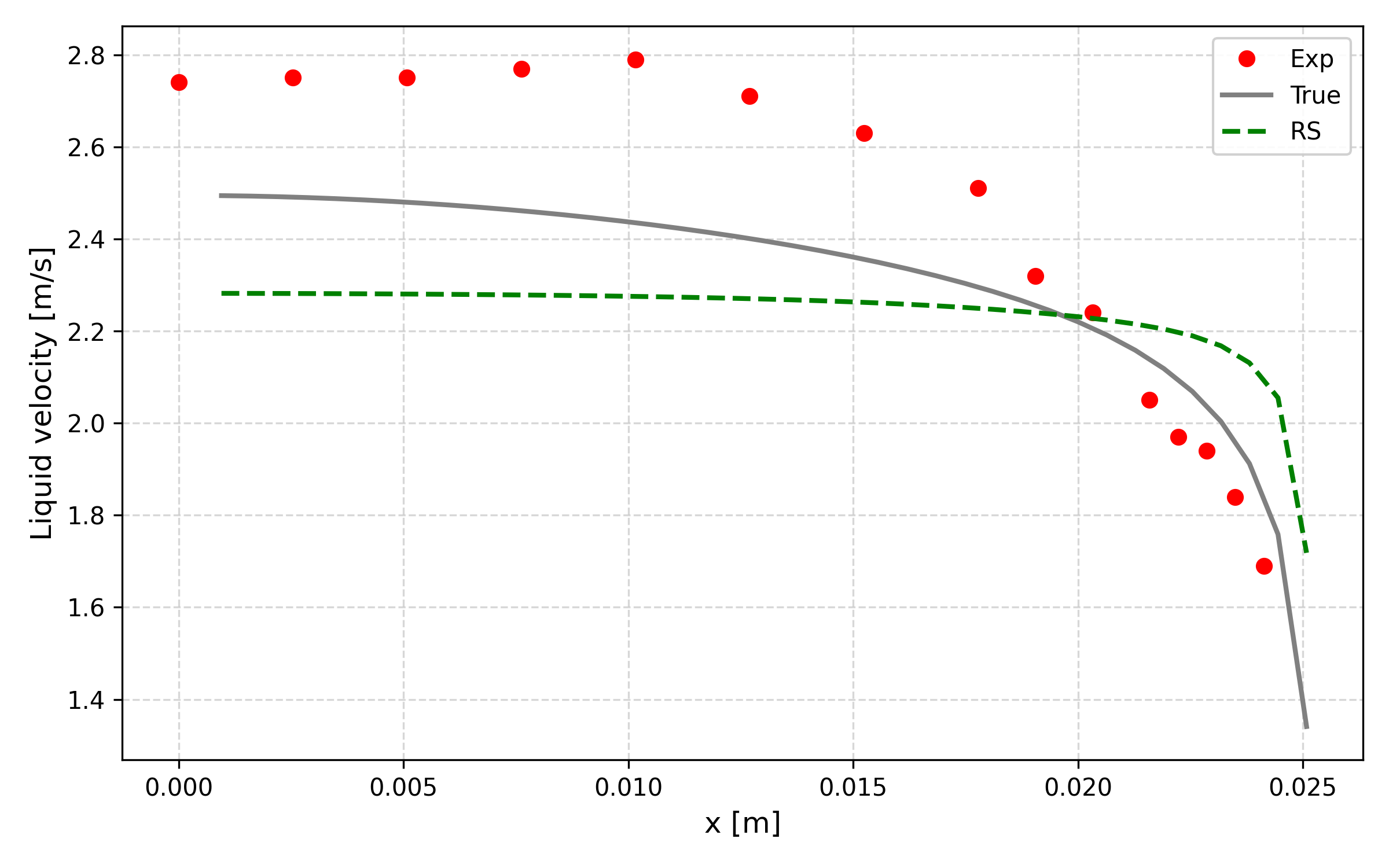}
        \caption{Liquid velocity}
        \label{fig:32_liqVel}
    \end{subfigure}
    \end{flushleft}

    \caption{Validation plots for a case example with high \textit{Regret}: Experiment ID 11/Case ID 97 with ${Regret}_c^{\text{RS}}=0.15$ ($p_c^{\text{RS}} = 0.65$, $p_c^\star = 0.80$). The recommendation via hybrid recommender system (RS, green dashed line) and the true best item (gray solid line) compared to the experimental validation data (red points).}
    \label{fig:high_regret}
\end{figure}

\begin{figure}
    \centering

    \begin{flushleft}
    \begin{subfigure}[b]{0.45\textwidth}
        \includegraphics[width=\textwidth]{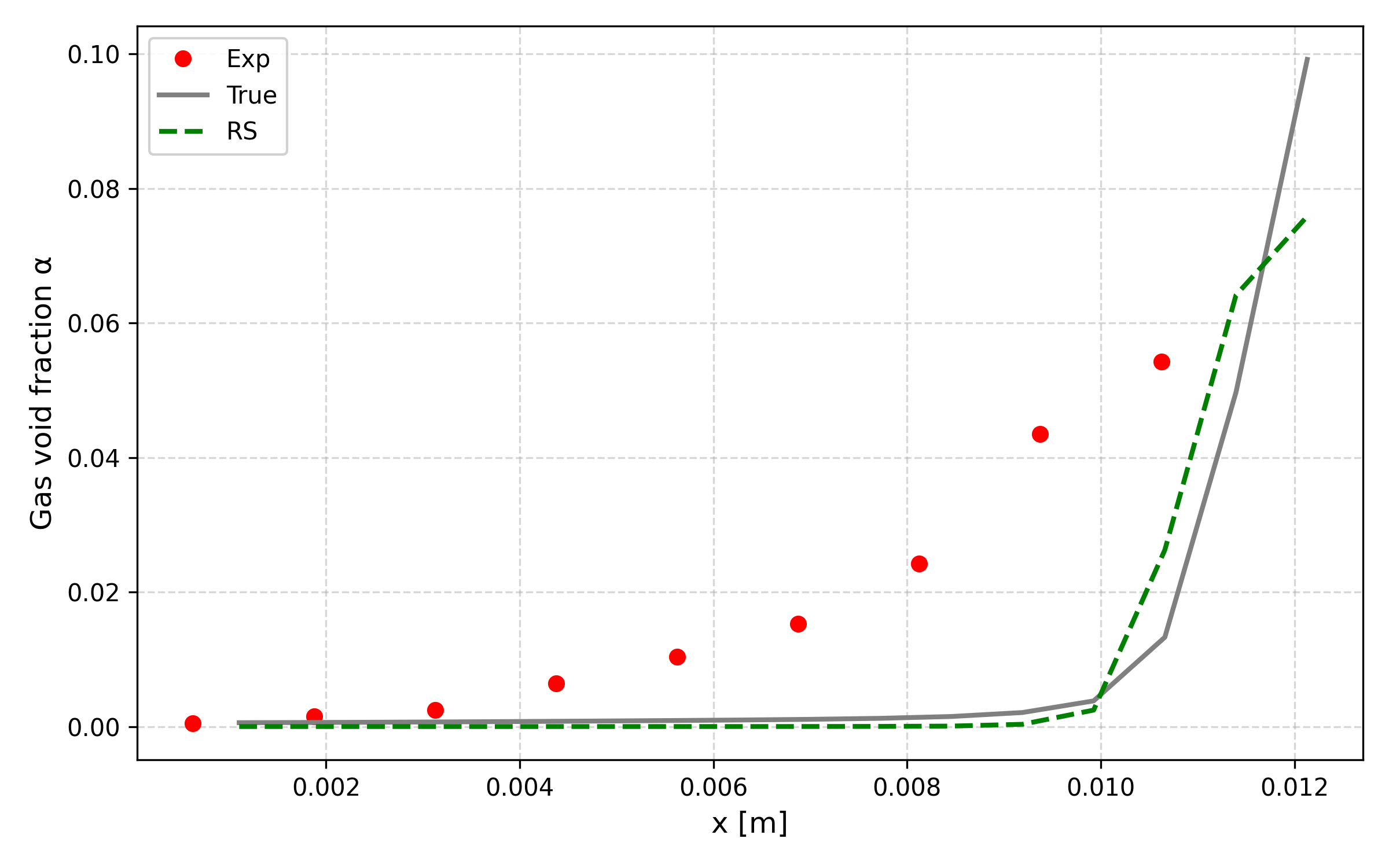}
        \caption{Gas void fraction}
        \label{fig:h21_alpha}
    \end{subfigure}
    \hspace{0.5cm}
    \begin{subfigure}[b]{0.45\textwidth}
        \includegraphics[width=\textwidth]{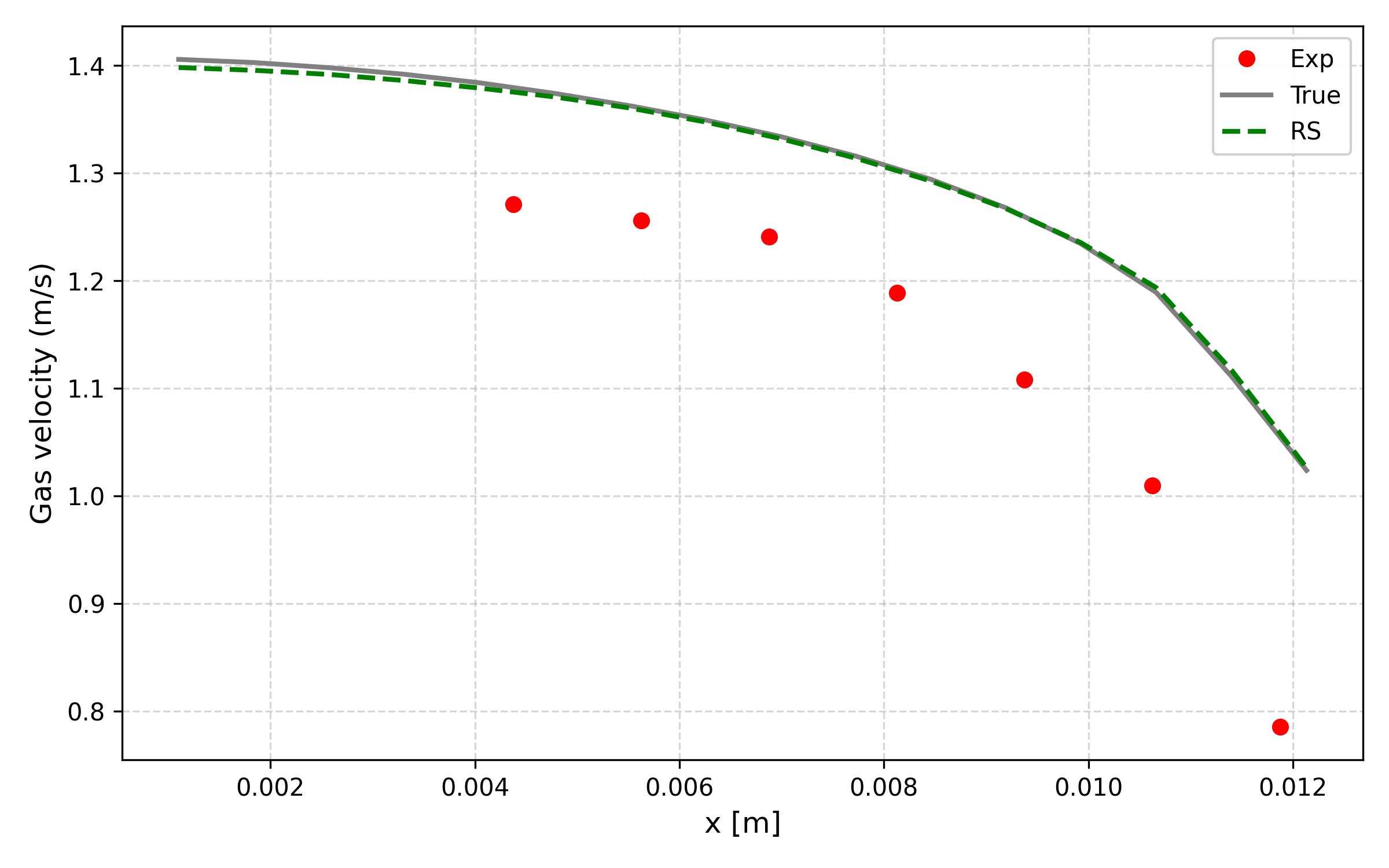}
        \caption{Gas velocity}
        \label{fig:h21_gasVel}
    \end{subfigure}
    \end{flushleft}

    \vspace{0.4cm}

    \begin{flushleft}
    \begin{subfigure}[b]{0.45\textwidth}
        \includegraphics[width=\textwidth]{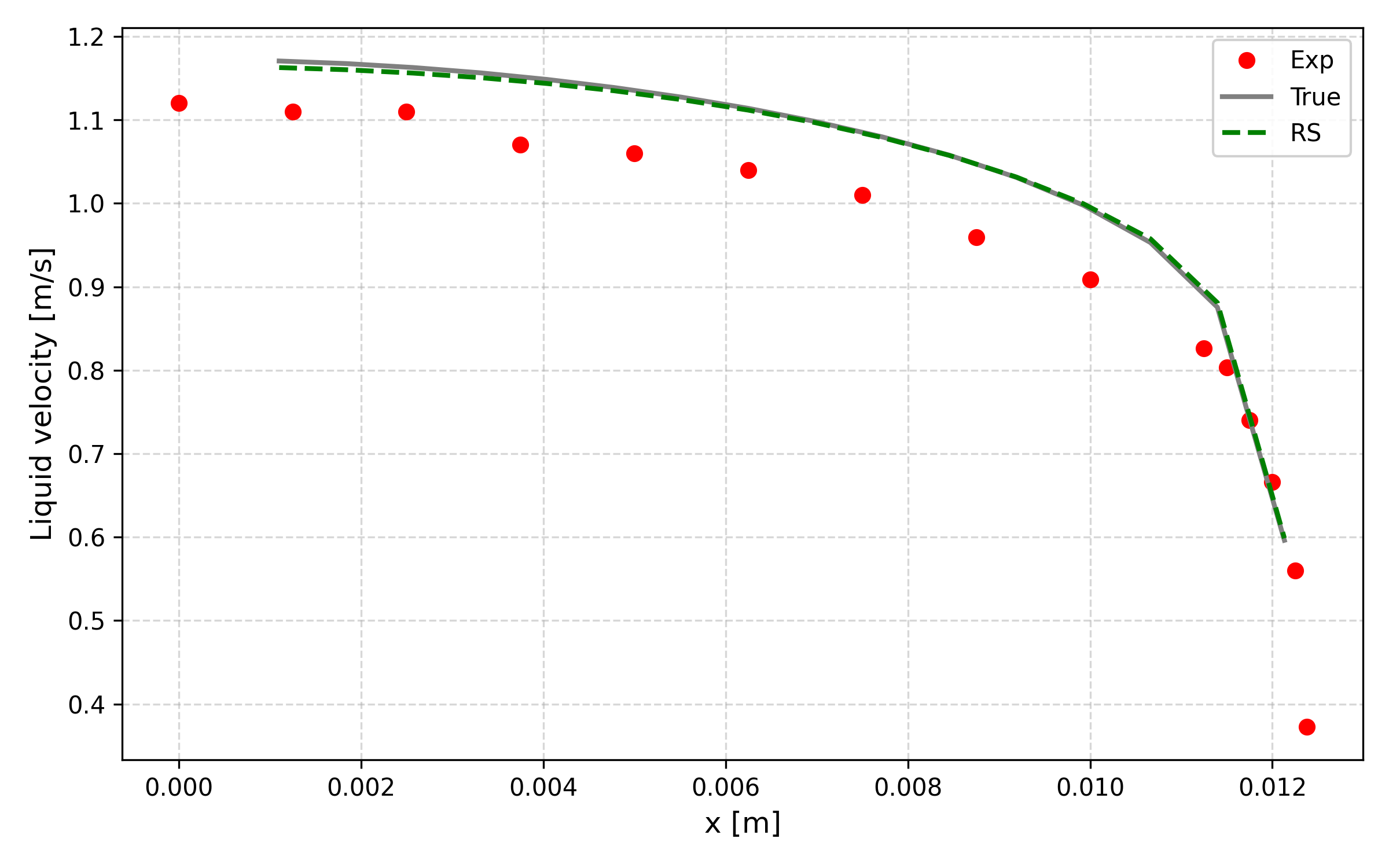}
        \caption{Liquid velocity}
        \label{fig:h21_liqVel}
    \end{subfigure}
    \hspace{0.5cm}
    \begin{subfigure}[b]{0.45\textwidth}
        \includegraphics[width=\textwidth]{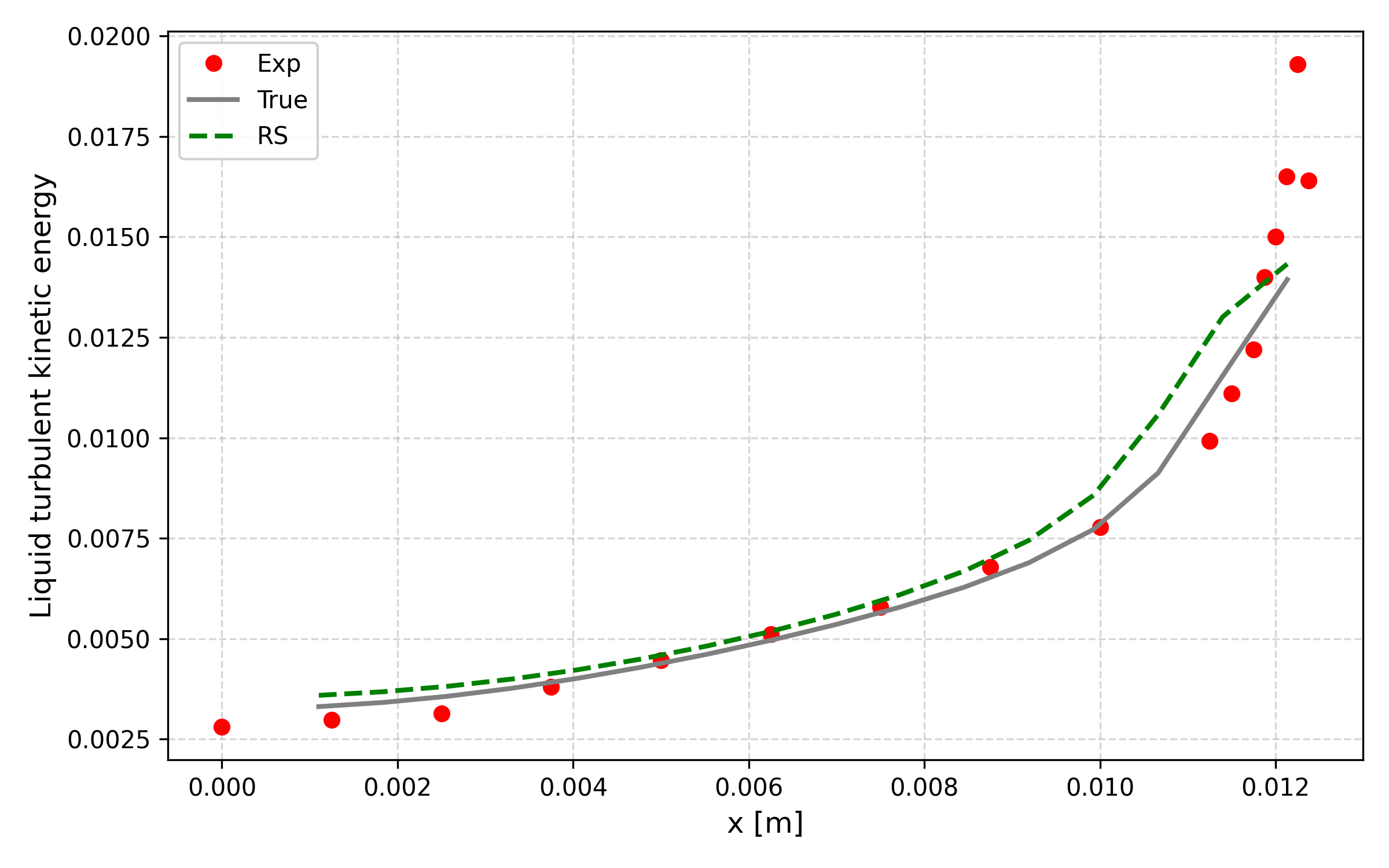}
        \caption{Liquid turbulent kinetic energy}
        \label{fig:h21_liqKe}
    \end{subfigure}
    \end{flushleft}

    \caption{Validation plots for a case example with low \textit{Regret}: Experiment ID 8/Case ID 49 with ${Regret}_c^{\text{RS}}=0.01$ ($p_c^{\text{RS}} = 0.81$ , $p_c^\star = 0.82$). The recommendation via hybrid recommender system (RS, green dashed line) and the true best item (gray solid line) compared to the experimental validation data (red points).}
    \label{fig:low_regret}
\end{figure}

Figures~\ref{fig:high_regret} and \ref{fig:low_regret} present representative high- and low-\textit{Regret} cases, illustrated by validation plots that compare the recommended closure model with the true best-performing item. For these illustrations, the item most frequently recommended by the recommender model across 100 independent scenarios at a sparsity level of $s = 0.75$ is shown, where each scenario corresponds to a different sparsified version of the performance matrix. Although quantitative results are generally reported as averages over these 100 scenarios, the \textit{Regret} values shown here are computed for the specific recommended items in order to ensure consistency with the displayed validation plots.

The examples demonstrate that large \textit{Regret} values are associated with qualitatively different flow predictions, such as a tendency towards an unphysical over-prediction of the gas void fraction where only a moderate wall peak profile is expected in Figure \ref{fig:32_alpha}. By contrast, low \textit{Regret} values correspond to near-optimal simulation results closely reproducing the experimentally observed flow structure in Figures \ref{fig:h21_alpha}-\ref{fig:h21_liqVel}.

\section{Conclusions}
CFD practitioners across a wide range of application domains are repeatedly confronted with the challenge of closure model selection for new simulation scenarios. In particular, the widely adopted Eulerian–Eulerian framework offers a large and continually expanding portfolio of closure models and associated parameters, driven by ongoing developments aimed at increasingly complex flow regimes. In the absence of a broadly accepted, universally applicable model set, model selection continues to rely on case-specific analyses and practitioner experience, often leading to extensive trial-and-error studies.

The work presented in this paper builds on the idea of applying recommender system methodologies to the long-standing challenge of closure model selection in multiphase CFD. By reframing closure model selection as an information-filtering problem, CFD cases are interpreted as users and closure model combinations as items, with case–item interactions quantified through a physically motivated performance metric. We propose a hybrid recommender system that integrates content-based similarity derived from case metadata with collaborative filtering applied to sparse performance data. The developed recommender model allows near-optimal closure models to be predicted for entirely new, unseen CFD cases based solely on their descriptive features. To our knowledge, this represents the very first cold-start recommender specifically developed for closure model selection in multiphase CFD.

The main contributions of this work can be summarised as:
(i) formulation of closure model selection as a cold-start recommendation problem in scientific computing,
(ii) development of a hybrid metadata-informed collaborative recommender framework,
(iii) risk-aware evaluation under experiment-level nested cross-validation.

The methodology was tested using a data set comprising 136 CFD cases of bubbly flow and 100 different model combinations for all lateral force combinations. Automated processing and evaluation of the result accuracy enabled us to generate a performance matrix of the case-model interactions as the data basis for our study. Based on this ground-truth matrix, different levels of sparsity were explored. After defining item relevance and selecting suitable evaluation metrics, we employed nested cross-validation to assess the generalisation capability of the hybrid recommender system, i.e. its ability to predict suitable models for entirely new CFD experiments.

Model recommendations were compared with a data-driven popularity baseline, and a reference model designed to be universally applicable for bubbly flows. The results demonstrate that the proposed $k$-NN-based hybrid recommender is capable of exploiting historical simulation data to provide meaningful model recommendations. Across a broad set of validation experiments, the recommender consistently outperformed a popularity baseline for all the investigated sparsity levels. In terms of \textit{Regret}, the recommender reduced performance loss relative to popularity-based selection, indicating an improved ability to identify closure models that are closer to optimal for a given case. An ablation analysis further demonstrated that the observed performance gains cannot be explained by matrix completion alone, but instead result from combining collaborative inference with neighbourhood modelling, highlighting the importance of leveraging structured case metadata for robust cold-start recommendation in scientific domains.

Comparison with an expert-designed reference closure model highlights an important complementary role of data-driven recommendation. While the reference model exhibits robust, sparsity-independent behaviour and moderate regret values --- reflecting its conservative and case-agnostic design --- it lacks adaptivity to case-specific flow conditions. The recommender system, by contrast, leverages similarities across cases to tailor recommendations to individual scenarios, thereby bridging the gap between universal modelling assumptions and fully case-specific optimisation.

Several limitations of the present study should be acknowledged. First, the closure model combinations considered represent only a subset of the full closure model design space; therefore, results should not be interpreted to form general conclusions regarding model performance. Second, numerical stability effects were treated implicitly by assigning zero performance to failed or oscillatory simulations, conflating physical inaccuracy and numerical robustness. A more explicit separation of these aspects, for example through multi-objective performance representations, constitutes an important direction for future work. Third, the results presented here assume that the entire list of case features is known as an input into the prediction. Finally, as with all similarity-based approaches, the recommender may fail for cases that are poorly represented in the available data or that lie outside the span of the feature space. Performance may degrade for extrapolated scenarios (e.g., new flow regimes or geometries not represented in the database), which motivates future work on uncertainty-aware or out-of-distribution detection mechanisms. 

Despite these limitations, the present study demonstrates the feasibility and potential of recommender systems as a decision-support tool for multiphase CFD model selection. Rather than replacing physical insight or expert judgement, the proposed approach provides a systematic mechanism for leveraging accumulated simulation experience at a scale that exceeds human memory and intuition. Beyond the specific application to multiphase CFD, this work illustrates how recommender system methodology can support decision-making in scientific domains characterised by expensive evaluations, structured metadata, and sparse historical observations. Many areas of computational science, where model configuration spaces are combinatorial, evaluation is expensive, and prior evidence is sparse, face similar challenges when selecting among competing model configurations. By demonstrating that hybrid, metadata-informed recommendation can generalise to entirely new physical scenarios, this study highlights the broader potential of AI-driven decision-support systems as a scalable mechanism for reusing accumulated simulation knowledge across complex scientific workflows.

Building on the evaluation framework established in this study, future work will explore alternative recommender system candidates for feature-based CFD model selection, such as inductive matrix factorisation \citep{ledent2021orthogonal}. Several promising areas for future work emerge from the presented findings. Priority will be given to expanding the closure model space, and to incorporating uncertainty estimates as well as stability-aware metrics. Further work will focus on simplifying the case feature input by an automated extraction of case metadata from the simulation setup files in the validation data base (see the work of \citep{Fan2026}), which will be enhanced by standardised and sufficiently detailed case descriptions. 

Collectively, these developments aim to enable more robust, transparent, and uncertainty-aware closure model selection in industrial CFD practice. Ultimately, the proposed framework illustrates how accumulated simulation evidence can be transformed from a passive archive into an active, data-driven decision-support system for scientific model selection.

\printcredits

\section*{Data availability}
Data will be made available on request.

\section*{Declaration of competing interests}

The authors declare that they have no known competing financial interests or personal relationships that could have appeared to influence the work reported in this paper.

\section*{Declaration of generative AI and AI-assisted technologies in the manuscript preparation process}

During the preparation of this work the authors used ChatGPT-5.2 in order to check the language and improve the readability of some paragraphs. After using this tool/service, the authors reviewed and edited the content as needed and take full responsibility for the content of the published article.

\appendix
\section{Appendix}

\subsection*{Nested Cross-Validation Procedure}
\label{app:cv_details}
This appendix provides a detailed description of the nested cross-validation procedure used to evaluate the recommender model, complementing the overview given in Algorithm~\ref{alg:nested_cv}.

For each sparsity level ($s = 0.25,\; 0.50,\; 0.75,\; 0.90$) of the ground-truth matrix $R$, we randomly remove entries to generate 100 random sparse realisations $R_s$. For each sparse matrix and each of the 17 CFD test experiments, the following steps are performed:

\begin{enumerate}
\renewcommand{\labelenumi}{\roman{enumi}.}
\item Remove the entire test experiment, including all of its sub-cases, from $R_s$ to form a sparse validation matrix $S_s$ (complete$-$test) (from which we can extract the popular item from observed entries).
\item Perform k-fold cross-validation on the remaining 16 experiments in $S_s$ to identify the optimal $k$-NN hyperparameters (see k-fold CV below).
\item Apply the matrix completion method (using \textit{gcimpute}) to estimate validation matrix $\tilde{S}_s$ (after which we can extract the popular item from observed and imputed entries).
\item Generate $k$-NN-based predictions for each test case by averaging observed and imputed performance values across its $k$ nearest neighbours.
\item Evaluate the ranking performance using $RR@k$ at the case level and $RR@k^{(\mathrm{exp})}$ at the experiment level.
\end{enumerate}

Averaging $RR@k^{(\mathrm{exp})}$ over the 17 test experiments yields the overall test metric $MRR@3_{\mathrm{test}}$.

\vspace{0.5cm}
For each test experiment, fair comparison among $k$-NN configurations is ensured through an inner k-fold cross-validation loop. The evaluated distance metrics include Euclidean, cosine, and Gower distances, while the number of neighbours is varied as $k \in \{1, 2, 3, 5, 10, 15, 20, 30, 50\}$, resulting in 21 distinct $k$-NN parameter configurations that were explored.

For each of the 16 validation experiments in $S_s$ (complete$-$test), the following procedure is applied:

\begin{enumerate}
\renewcommand{\labelenumi}{\roman{enumi}.}
\item Remove the validation experiment and all its sub-cases from $S_s$ to obtain a sparse training matrix $T_s$ (complete$-$test$-$validation).
\item Apply the matrix completion method (\textit{gcimpute}) to estimate the training matrix $\tilde{T}_s$.
\item For each of the 21 $k$-NN parameter configurations:
\begin{enumerate}
    \item Identify the $k$ nearest neighbours for each case of the validation experiment based on its specific case features.
    \item Predict validation-case performances by averaging over similar cases using the 100 imputed training matrices.
    \item Compute ranking metrics (e.g. $RR@3$) at the sub-case level.
    \item Average metrics across cases belonging to the same validation experiment to obtain $RR@3^{(\mathrm{exp})}$.
\end{enumerate}

\item Average results across all 16 validation experiments to compute $MRR@3_{\mathrm{val}}$.
\item Select the optimal $k$-NN parameter configuration based on the $MRR@3_{\mathrm{val}}$.
\end{enumerate}

This nested cross-validation procedure ensures that the data of each test experiment was not seen during matrix completion nor the $k$-NN hyperparameter selection, thereby providing an unbiased estimate of model performance on unseen CFD experiments. 

\newpage
\subsection*{Detailed cross-validation results}
\label{app:cv_results}

\begin{table}
\centering
\caption{Best $k$-NN parameters for the individual test experiments for $s=0.75$ with mean and 95\% confidence intervals (CI).}
\label{tab:best_knn_ci}
\resizebox{\textwidth}{!}{%
\begin{tabular}{ccccccc}
\toprule
\textbf{Exp.ID} & \textbf{Metric} & \textbf{k} & \textbf{RR@3 (val)} & \textbf{RR@3 (test)} & \textbf{RR@1 (val)} & \textbf{RR@1 (test)} \\
&  &  & mean & mean [CI] & mean & mean [CI] \\
\midrule
1  & cosine & 10 & 0.496 & 0.437 [0.403, 0.472] & 0.389 & 0.325 [0.287, 0.363] \\
2  & cosine & 10 & 0.491 & 0.312 [0.271, 0.352] & 0.383 & 0.235 [0.193, 0.277] \\
3  & euclidean & 10 & 0.515 & 0.273 [0.255, 0.291] & 0.400 & 0.189 [0.171, 0.208] \\
4  & cosine & 10 & 0.472 & 0.813 [0.785, 0.841] & 0.365 & 0.677 [0.631, 0.724] \\
5  & cosine & 10 & 0.481 & 0.692 [0.655, 0.729] & 0.377 & 0.560 [0.511, 0.609] \\
6  & cosine & 10 & 0.480 & 0.641 [0.606, 0.675] & 0.373 & 0.496 [0.452, 0.540] \\
7  & cosine & 10 & 0.504 & 0.416 [0.382, 0.449] & 0.394 & 0.292 [0.255, 0.328] \\
8  & cosine & 10 & 0.467 & 0.835 [0.807, 0.862] & 0.358 & 0.725 [0.681, 0.769] \\
9  & euclidean & 50 & 0.491 & 0.138 [0.110, 0.166] & 0.378 & 0.072 [0.047, 0.098] \\
10  & cosine & 10 & 0.462 & 0.594 [0.581, 0.607] & 0.360 & 0.476 [0.461, 0.491] \\
11  & gower & 50 & 0.467 & 0.358 [0.324, 0.392] & 0.361 & 0.273 [0.238, 0.309] \\
12  & cosine & 10 & 0.482 & 0.683 [0.612, 0.755] & 0.375 & 0.520 [0.420, 0.620] \\
13  & cosine & 10 & 0.481 & 0.922 [0.884, 0.959] & 0.373 & 0.850 [0.779, 0.921] \\
14  & cosine & 10 & 0.487 & 0.399 [0.358, 0.441] & 0.378 & 0.287 [0.243, 0.332] \\
15  & cosine & 10 & 0.475 & 0.652 [0.595, 0.710] & 0.370 & 0.545 [0.475, 0.615] \\
16  & cosine & 10 & 0.449 & 0.568 [0.550, 0.586] & 0.332 & 0.485 [0.465, 0.505] \\
17  & cosine & 10 & 0.477 & 0.987 [0.977, 0.996] & 0.368 & 0.973 [0.955, 0.992] \\
\bottomrule
\end{tabular}
}
\end{table}

\subsection*{Per experiment sparsity plots}

\begin{figure}
  \centering
  \includegraphics[width=0.8\textwidth]{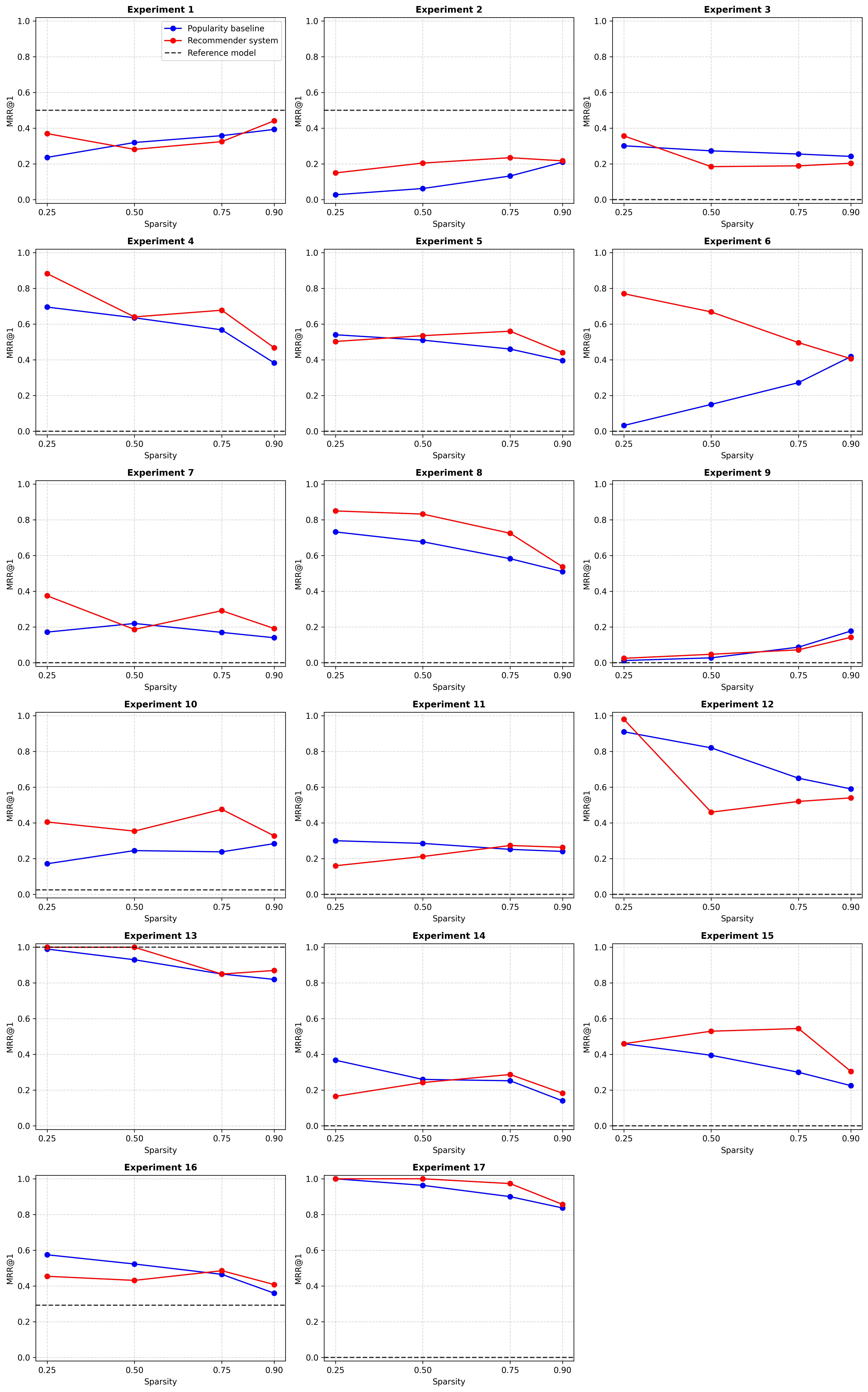}
  \caption{\textit{MRR@1} results across different sparsity levels plotted for each individual CFD experiment. Comparison of the popularity baseline, the recommender prediction and the reference item.}
  \label{fig:sparsity_mrr1PerExp}
\end{figure}

\begin{figure}
  \centering
  \includegraphics[width=0.8\textwidth]{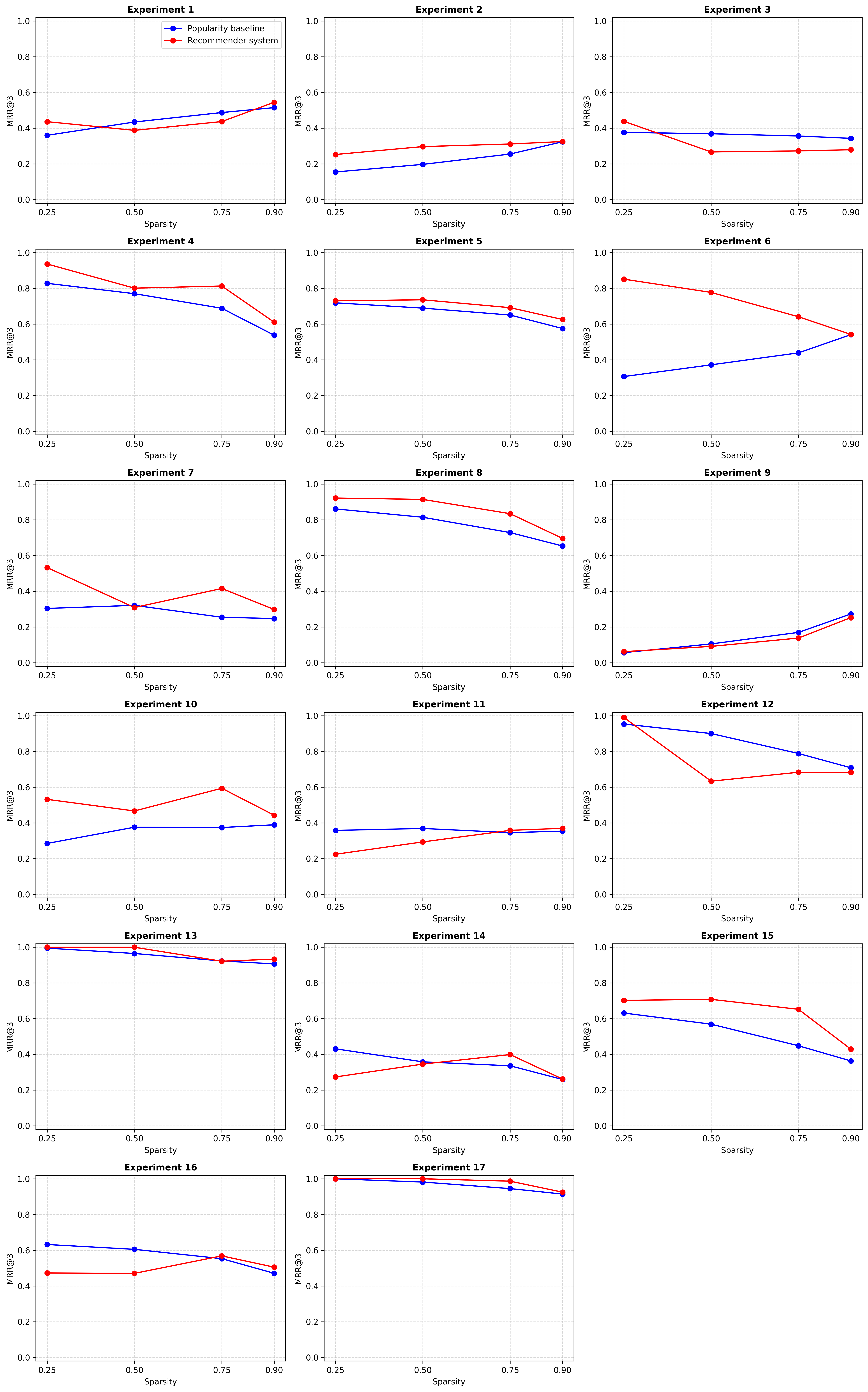}
  \caption{\textit{MRR@3} results across different sparsity levels plotted for each individual CFD experiment. Comparison of the popularity baseline, the recommender prediction and the reference item.}
  \label{fig:sparsity_mrr3PerExp}
\end{figure}

\begin{figure}
  \centering
  \includegraphics[width=0.8\textwidth]{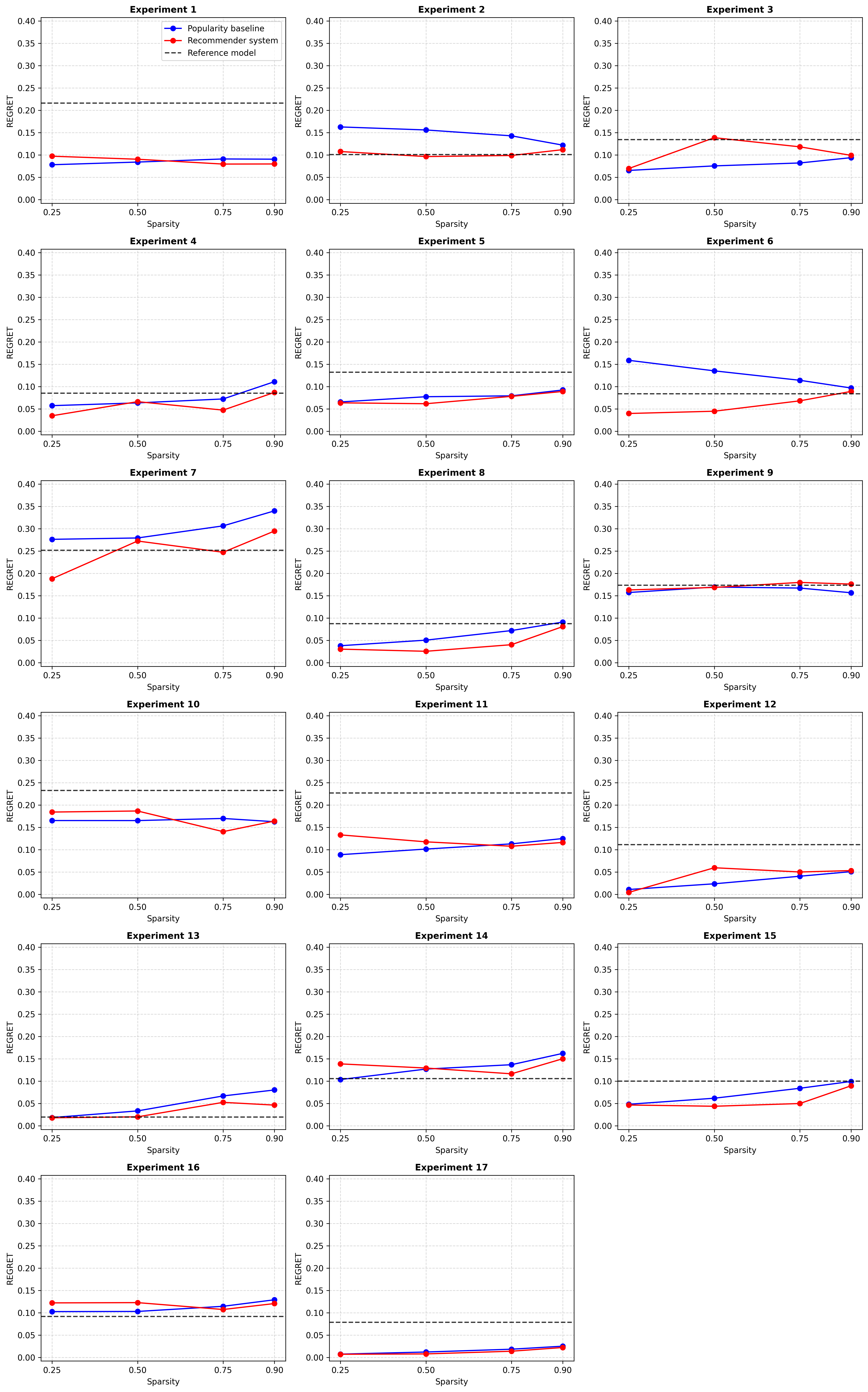}
  \caption{\textit{Regret} results across different sparsity levels plotted for each individual CFD experiment. Comparison of the popularity baseline, the recommender prediction and the reference item.}
  \label{fig:sparsity_regretPerExp}
\end{figure}

\bibliographystyle{cas-model2-names}

\bibliography{literature}

@book{Yeoh2019,
  author    = {Guan Heng Yeoh and Jiyuan Tu},
  title     = {Computational Techniques for Multiphase Flows},
  edition   = {2},
  publisher = {Butterworth-Heinemann},
  address   = {Oxford},
  year      = {2019},
  isbn      = {9780081024546},
}

@book{Ishii2011,
  author    = {Mamoru Ishii and Takashi Hibiki},
  title     = {Thermo-Fluid Dynamics of Two-Phase Flow},
  publisher = {Springer},
  address   = {New York},
  year      = {2011},
  edition   = {2},
  doi       = {10.1007/978-1-4419-7985-8},
  isbn      = {978-1-4419-7984-1}
}

@article{Fan2026,
  author       = {E. Fan and Kang Hu and Zhuowen Wu and Jiangyang Ge and Jiawei Miao and Yuzhi Zhang and He Sun and Weizong Wang and Tianhan Zhang},
  title        = {ChatCFD: A Large Language Model-Driven Agent for End-to-End Computational Fluid Dynamics Automation with Structured Knowledge and Reasoning},
  journal      = {Advanced Intelligent Discovery},
  year         = {2026},
  doi          = {10.1002/aidi.202500174},
  url          = {https://doi.org/10.1002/aidi.202500174},
}

@article{Pandey2025,
  title        = {OpenFOAMGPT: a RAG-Augmented LLM Agent for OpenFOAM-Based Computational Fluid Dynamics},
  author       = {Pandey, Sandeep and Xu, Ran and Wang, Wenkang and Chu, Xu},
  year         = {2025},
  journal      = {Physics of Fluids},
  volume       = {37},
  number       = {3}
}

@article{Dong2025,
  title={Fine-tuning a large language model for automating computational fluid dynamics simulations},
  author={Zhehao Dong and Zhen Lu and Yue Yang},
  journal={Theoretical and Applied Mechanics Letters},
  volume={15},
  number={3},
  year={2025}
}

@incollection{Ricci2022,
  author       = {Francesco Ricci and Lior Rokach and Bracha Shapira},
  title        = {Recommender Systems: Techniques, Applications, and Challenges},
  booktitle    = {Recommender Systems Handbook},
  editors       = {Francesco Ricci and Lior Rokach and Bracha Shapira},
  publisher    = {Springer, New York, NY},
  year         = {2022},
  edition      = {3rd},
  doi          = {10.1007/978-1-0716-2197-4_1}
}

@inproceedings{spivsak2025segment,
  title={Segment-Aware Analytics for Real-Time Editorial Support in Media Groups},
  author={Spi{\v{s}}{\'a}k, Martin and Alves, Rodrigo and Kelleher, Thomas and Sheppard, Jason and Fiedler, Ond{\v{r}}ej and Kosovrasti, Ergys and Van{\v{c}}ura, Vojt{\v{e}}ch and Kasalick{\`y}, Petr and Kord{\'\i}k, Pavel},
  booktitle    = {INRA 2025: 13th International Workshop on News Recommendation and Analytics co-located with the 19th ACM Conference on Recommender Systems, Prague, Czech Republic, September 22--26, 2025},
  pages        = {1--8},
  publisher    = {CEUR Workshop Proceedings},
  series       = {CEUR-WS.org},
  volume       = {4056},
  url          = {https://ceur-ws.org/Vol-4056/short1.pdf},
  year={2025}
}

@article{ledent2021orthogonal,
  title={Orthogonal inductive matrix completion},
  author={Ledent, Antoine and Alves, Rodrigo and Kloft, Marius},
  journal={IEEE Transactions on Neural Networks and Learning Systems},
  volume={34},
  number={5},
  pages={2259--2270},
  year={2021},
  publisher={IEEE}
}

@article{johnson2014logistic,
  title={Logistic matrix factorization for implicit feedback data},
  author={Johnson, Christopher C and others},
  journal={Advances in Neural Information Processing Systems},
  volume={27},
  number={78},
  pages={1--9},
  year={2014},
  publisher={Montr{\'e}al, Canada}
}

@article{schedl2019deep,
  title={Deep learning in music recommendation systems},
  author={Schedl, Markus},
  journal={Frontiers in Applied Mathematics and Statistics},
  volume={5},
  pages={44},
  year={2019},
  publisher={Frontiers Media SA}
}

@article{alves2024regionalization,
  title={Regionalization-Based Collaborative Filtering: Harnessing Geographical Information in Recommenders},
  author={Alves, Rodrigo},
  journal={ACM Transactions on Spatial Algorithms and Systems},
  volume={10},
  number={2},
  pages={1--23},
  year={2024},
  publisher={ACM New York, NY}
}

@article{mazumder2010spectral,
  title={Spectral regularization algorithms for learning large incomplete matrices},
  author={Mazumder, Rahul and Hastie, Trevor and Tibshirani, Robert},
  journal={The Journal of Machine Learning Research},
  volume={11},
  pages={2287--2322},
  year={2010},
  publisher={JMLR. org}
}

@article{Huang2023,
  author = {X. Huang and T. Chyczewski and Z. Xia and R. Kunz and X. Yang}, 
  title = {Distilling experience into a physically interpretable recommender system for computational model selection},
  journal = {Scientific Reports},
  volume = {13},
  pages = {2225},
  year = {2023},
  doi = {https://doi.org/10.1038/s41598-023-27426-5},
}

@article{Misir2017,
  title     = {Alors: An algorithm recommender system},
  author    = {Mısır, Mustafa and Sebag, Michèle},
  journal   = {Artificial Intelligence},
  volume    = {244},
  pages     = {291--314},
  year      = {2017},
  publisher = {Elsevier},
  doi       = {10.1016/j.artint.2016.12.001}
}

@book{Falk2019,
  author    = {Kim Falk},
  title     = {Practical Recommender Systems},
  publisher = {Manning Publications / Simon \& Schuster},
  year      = {2019},
  isbn      = {9781617292705},
  pages     = {432},
  address   = {New York, NY},
}

@article{Zhao2024,
  title   = {gcimpute: A Package for Missing Data Imputation},
  author  = {Zhao, Yuxuan and Udell, Madeleine},
  journal = {Journal of Statistical Software},
  volume  = {108},
  number  = {4},
  year    = {2024},
  month   = {February},
  doi     = {10.18637/jss.v108.i04}
}

@article{Haensch2025,
  author = {S. H{\"a}nsch and A. Sajdoková and A. Rabau and V. Rybář and R. Alves and P. Kordík}, 
  title = {Data-driven closure model selection for multiphase CFD via matrix completion},
  journal = {AI Thermal Fluids},
  volume = {4},
  pages = {100019},
  year = {2025},
  doi = {https://doi.org/10.1016/j.aitf.2025.100019},
}

@book{Goodfellow2016,
  title={Deep Learning},
  author={Goodfellow, Ian and Bengio, Yoshua and Courville, Aaron},
  year={2016},
  publisher={MIT Press}
}

@article{Zhang2019,
  title={Deep learning based recommender system: A survey and new perspectives},
  author={Zhang, Shuai and others},
  journal={ACM Computing Surveys},
  year={2019}
}

@article{Lucas2016a,
  author       = {Lucas, D. and Rzehak, R. and Krepper, E. and Ziegenhein, Th. and Liao, Y. and Kriebitzsch, S. and Apanasevich, P.},
  title        = {A Strategy for the Qualification of Multi‑Fluid Approaches for Nuclear Reactor Safety},
  journal      = {Nuclear Engineering and Design},
  volume       = {299},
  pages        = {2--11},
  year         = {2016},
  doi          = {10.1016/j.nucengdes.2015.07.007},
}

@article{Rzehak2017,
  author       = {R. Rzehak and T. Ziegenhein and S. Kriebitzsch and E. Krepper and D. Lucas},
  title        = {Unified modeling of bubbly flows in pipes, bubble columns, and airlift columns},
  journal      = {Chemical Engineering Science},
  year         = {2017},
  volume       = {157},
  pages        = {147--158},
  doi          = {10.1016/j.ces.2016.04.056},
  url          = {https://www.sciencedirect.com/science/article/pii/S0009250916302226},
  publisher    = {Elsevier}
}

@article{Liao2020,
title = "Eulerian-{E}ulerian two-fluid model for laminar bubbly pipe flows: Validation of the baseline model",
journal = "Computers \& Fluids",
volume = "202",
pages = "104496",
year = "2020",
issn = "0045-7930",
doi = "https://doi.org/10.1016/j.compfluid.2020.104496",
url = "http://www.sciencedirect.com/science/article/pii/S0045793020300694",
author = "Yixiang Liao and Kartik Upadhyay and Fabian Schlegel",
}

@article{Colombo2021,
  author = {M. Colombo and R. Rzehak and M. Fairweather and Y. Liao and D. Lucas}, 
  title = {Benchmarking of computational fluid dynamic models for bubbly flows},
  journal = {Nuclear Engineering and Design},
  volume = {375},
  pages = {111075},
  year = {2021},
  doi = {https://doi.org/10.1016/j.nucengdes.2021.111075},
}

@article{Besagni2023,
  author    = {Giorgio Besagni and Nicolò Varallo and Riccardo Mereu},
  title     = {Computational Fluid Dynamics Modelling of Two-Phase Bubble Columns: A Comprehensive Review},
  journal   = {Fluids},
  year      = {2023},
  volume    = {8},
  number    = {3},
  pages     = {91},
  doi       = {10.3390/fluids8030091},
  url       = {https://www.mdpi.com/2311-5521/8/3/91},
  publisher = {MDPI},
  note      = {Open Access under CC BY 4.0},
}

@article{Gray2021,
  title     = {A comparative study of closure relations for {CFD} modelling of bubbly flow in a vertical pipe},
  author    = {Gray, Geoffrey S. and Ormiston, Scott J.},
  journal   = {Open J. Fluid Dyn.},
  publisher = {Scientific Research Publishing, Inc.},
  volume    =  {11},
  number    =  {02},
  pages     = {98--134},
  year      =  {2021}
}

@article{Adzaklo2025,
  author = {Simon Yao Adzaklo and Kwame Sarkodie and Emmanuel Agyei and Stephen Yamoah and Nana Yaw Asiedu},
  title = {Investigation of drag models for simulation of gas-liquid two-phase flow systems},
  journal = {Scientific Reports},
  year = {2025},
  volume = {15},
  number = {1},
  pages = {39440},
  doi = {10.1038/s41598-025-23038-3},
  url = {https://www.nature.com/articles/s41598-025-23038-3},
  publisher = {Springer Nature}
}

@article{Garcia-Villalba2025,
  title        = {Numerical methods for multiphase flows},
  author       = {Garcia‐Villalba, Manuel and Colonius, Tim and Desjardins, Olivier and Lucas, Dirk and Mani, Ali and Marchisio, Daniele and Matar, Omar K. and Picano, Francesco and Zaleski, Stéphane},
  journal      = {International Journal of Multiphase Flow},
  volume       = {191},
  pages        = {105285},
  year         = {2025},
  doi          = {10.1016/j.ijmultiphaseflow.2025.105285},
  url          = {https://www.sciencedirect.com/science/article/pii/S0301932225001636}
}

@article{Haensch2021,
  author  = {S. H{\"a}nsch and I. Evdokimov and F. Schlegel and D. Lucas},
  journal = {Chemical Engineering Science},
  title   = {A workflow for the sustainable development of closure models for bubbly flows},
  year    = {2021},
  pages   = {116807},
  volume  = {244},
}

@article{Lehnigk2023,
  author  = {R. Lehnigk and M. Bruschewski and T. Huste and D. Lucas and and M. Rehm and F. Schlegel},
  journal = {Kerntechnik},
  title   = {Sustainable development of simulation setups and addons for OpenFOAM for nuclear reactor safety research},
  year    = {2023},
  pages   = {pp. 131-140},
  volume  = {88},
  number  = {2},
  doi     = {10.1515/kern-2022-0107},
}

@misc{HZDRcode,
	author = {F. Schlegel and K.G. Bilde and M. Draw and I. Evdokimov and S. H{\"a}nsch and V.V. Kamble and H. Khan and B. Krull and R. Lehnigk and J. Li and H. Lyu. and R. Meller and G. Petelin and S.P. Kota and M. Tekav\v{c}i\v{c}},
	title = {{M}ultiphase {C}ode {R}epository by {HZDR} for {O}pen{FOAM} {F}oundation {S}oftware},
	howpublished = {\url{https://rodare.hzdr.de/record/3284}},
	doi = {10.14278/rodare.3284},
	year = {2024}
}

@misc{HZDRpython,
	author = {F. Schlegel and C. Fombonne and S. H{\"a}nsch and B. Krull and R. Lehnigk and R. Meller},
	title = {{M}ultiphase {P}ython {R}epository by {HZDR}},
	howpublished = {\url{https://rodare.hzdr.de/record/3569}},
	doi = {10.14278/rodare.3569},
	year = {2025}
}

@misc{HZDRcases,
	author = {S. H{\"a}nsch and M. Draw and I. Evdokimov and H. Khan and B. Krull and R. Lehnigk and Y. Liao and H. Lyu, H. and R. Meller and F. Schlegel and M. Tekav\v{c}i\v{c}},
	title = {{M}ultiphase {C}ases {R}epository by {HZDR} for {O}pen{FOAM} {F}oundation {S}oftware},
	howpublished = {\url{https://rodare.hzdr.de/record/3862}},
	doi = {10.14278/rodare.3862},
	year = {2025}
}

@manual{snakemakeLibrary,
  title        = {Snakemake:  A framework for reproducible data analysis},
  author       = {{Snakemake Developers}},
  year         = {2025},
  url          = {https://snakemake.github.io/},
  note         = {Accessed: July 9, 2025}
}

@article{Pedregosa2011,
  title={Scikit-learn: Machine Learning in Python},
  author={Pedregosa, Fabian and Varoquaux, Ga{\"e}l and Gramfort, Alexandre and Michel, Vincent and Thirion, Bertrand and Grisel, Olivier and Blondel, Mathieu and Prettenhofer, Peter and Weiss, Ron and Dubourg, Vincent and Vanderplas, Jake and Passos, Alexandre and Cournapeau, David and Brucher, Matthieu and Perrot, Matthieu and Duchesnay, {\'E}douard},
  journal={Journal of Machine Learning Research},
  volume={12},
  pages={2825--2830},
  year={2011}
}

@article{IshiiZuber1979,
  author  = {Ishii, Mamoru and Zuber, Novak},
  title   = {Drag coefficient and relative velocity in bubbly, droplet or particulate flows},
  journal = {AIChE Journal},
  year    = {1979},
  volume  = {25},
  number  = {5},
  pages   = {843--855},
  month   = sep,
  doi     = {10.1002/aic.690250513},
  url     = {https://doi.org/10.1002/aic.690250513}
}

@book{Crowe2011,
author = "C.T. Crowe and J.D. Schwarzkopf and M. Sommerfeld, M. and Y. Tsuji",
title = "Multiphase Flows with Droplets and Particles",
editor = "Boca Raton, USA",
year = "2011",
}

@article{Hessenkemper2021,
  author  = {H. Hessenkemper and T. Ziegenhein and R. Rzehak and D. Lucas and A. Tomiyama},
  journal = {International Journal of Multiphase Flow},
  title   = {Lift force coefficient of ellipsoidal single bubbles in water},
  year    = {2021},
  pages   = {103587},
  volume  = {138}
}

@article{Legendre1998,
title = "The lift force on a spherical bubble in a viscous linear shear flow",
journal = "J. Fluid Mech.",
volume = "368",
pages = "81 - 126",
year = "1998",
author = "D. Legendre and J. Magnaudet",
}

@article{Moraga1999,
title = {Lateral forces on spheres in turbulent uniform shear flow},
journal = {International Journal of Multiphase Flow},
volume = {25},
number = {6 - 7},
pages = {1321 - 1372},
year = {1999},
author = {F. J. Moraga and F. J. Bonetto and R. T. Lahey},
}

@article{Tomiyama2002b,
  author  = {A. Tomiyama and H. Tamai and I. Zun and S. Hosokawa},
  journal = {Chemical Engineering Science},
  title   = {Transverse migration of single bubbles in simple shear flows},
  year    = {2002},
  pages   = {1849 – 1858},
  volume  = {57},
}

@article{Gosman1992,
title = "Multidimensional modeling of turbulent two-phase flows in stirred vessels",
journal = "AIChE Journal",
volume = "38",
number = "12",
pages = "1946-1956",
year = "1992",
author = "Gosman, A. D. and Lekakou, C. and Politis, S. and Issa, R. I. and Looney, M. K.",
}

@phdthesis{LopezDeBertodano1992,
  author       = "Lopez de Bertodano, M.", 
  title        = "Turbulent bubbly two-phase flow in a triangular duct",
  school       = "Rensselaer Polytechnic Institution",
  year         = "1992",
}

@inproceedings{Burns2004,
author = "A. Burns and T. Frank and I. Hamill and J.-M. Shi",
title = "The {F}avre Averaged Drag Model for Turbulent Dispersion in {E}ulerian Multi-Phase Flows",
year = "2004",
place = "Yokohama, Japan",
booktitle = "5th International Conference on Multiphase Flow",
}

@article{Antal1991,
title = "Analysis of phase distribution in fully developed laminar bubbly two-phase flow",
journal = "International Journal of Multiphase Flow",
volume = "17",
number ="5",
pages = "635-652",
year = "1991",
author = "Antal, S. P. and Lahey Jr, R.T. and Flaherty, J.E.",
}

@article{Tomiyama1998b,
title = "Struggle with computational bubble dynamics",
journal = "Multiphase Science and Technology",
volume ="10",
number ="4",
year ="1998",
pages ="369-405",
author ="Tomiyama, A.",
}

@inproceedings{Hosokawa2002,
author = "S. Hosokawa and A. Tomiyama and S. Misaki and T. Hamada",
title = "Lateral migration of single bubbles due to the presence of wall.",
year = "2002",
booktitle = "ASME 2002 Joint US-European Fluids Engineering Division Conference",
}

@inproceedings{Frank2005,
  author    = "Frank, T.",
  title     = "Advances in computational fluid dynamics ({CFD}) of 3-dimensional gas-liquid multiphase flows",
  year      = "2005",
  pages     = "1-18",
  place     = "Wiesbaden, Germany",
  booktitle = "NAFEMS Seminar: Simulation of Complex Flows (CFD)-Applications and Trends",
}

@article{Menter2003,
title = "Ten Years of Industrial Experience with the {SST} Turbulence Model",
journal = "Turbulence, Heat and Mass Transfer",
volume = "4",
pages = "625 - 632",
year = "2003",
author = "F.R. Menter and M. Kuntz and R. Langtry",
editor = "K. Hanjalic and Y. Nagano and M. Tummers, Begell House, Inc.",
}

@article{Ma2017,
title = "Direct numerical simulation-based {R}eynolds-averaged closure for bubble-induced turbulence",
journal = "Physical Review Fluids",
volume = "2",
year = "2017",
author = "T. Ma and C. Santarelli and T. Ziegenhein and D. Lucas and J. Fr{\"o}hlich",
}

@article{Liao2015,
title = "Baseline closure model for dispersed bubbly flow: Bubble coalescence and breakup",
journal = "Chemical Engineering Science",
volume = "122",
pages = "336 - 349",
year = "2015",
issn = "0009-2509",
doi = "https://doi.org/10.1016/j.ces.2014.09.042",
url = "http://www.sciencedirect.com/science/article/pii/S0009250914005466",
author = "Yixiang Liao and Roland Rzehak and Dirk Lucas and Eckhard Krepper",
}

@article{Sommer2023,
  author  = {Anna-Elisabeth Sommer and Mazen Draw and Lantian Wang and Jan Schmidtpeter and Hendrik Hessenkemper and Josefine Gatter and Haein Nam and Kerstin Eckert and Roland Rzehak},
  title   = {Hydrodynamics in a Bubble Column – Part 1: Two-Phase Flow},
  journal = {Chemical Engineering \& Technology},
  volume  = {46},
  number  = {9},
  pages   = {1763--1772},
  year    = {2023},
  doi     = {10.1002/ceat.202300130}
}

@article{NeumannKipping2020,
  author  = {Neumann-Kipping, Martin and Bieberle, André and Hampel, Uwe},
  title   = {Investigations on bubbly two-phase flow in a constricted vertical pipe},
  journal = {International Journal of Multiphase Flow},
  volume  = {130},
  pages   = {103340},
  year    = {2020},
  doi     = {10.1016/j.ijmultiphaseflow.2020.103340}
}

@article{Ziegenhein2019,
  author  = {Ziegenhein, Thomas and Lucas, Dirk},
  title   = {The critical bubble diameter of the lift force in technical and environmental, buoyancy-driven bubbly flows},
  journal = {International Journal of Multiphase Flow},
  volume  = {116},
  pages   = {26--38},
  year    = {2019},
  doi     = {10.1016/j.ijmultiphaseflow.2019.03.002}
}

@article{Kim2016,
  author  = {Kim, Minseok and Lee, Jae-Hun and Park, Hyungmin},
  title   = {Study of bubble-induced turbulence in upward laminar bubbly pipe flows measured with a two-phase particle image velocimetry},
  journal = {Experiments in Fluids},
  volume  = {57},
  number  = {12},
  pages   = {186},
  year    = {2016},
  doi     = {10.1007/s00348-016-2272-3}
}

@article{Hosokawa2013,
title = "Bubble-induced pseudo turbulence in laminar pipe flows",
journal = "International Journal of Heat and Fluid Flow",
volume = "40",
pages = "97 - 105",
year = "2013",
issn = "0142-727X",
doi = "https://doi.org/10.1016/j.ijheatfluidflow.2013.01.004",
url = "http://www.sciencedirect.com/science/article/pii/S0142727X13000088",
author = "S. Hosokawa and A. Tomiyama",
}

@article{Lucas2010,
  author  = {Lucas, Dirk and Beyer, Martin and Szalinski, Ludovic and Sch{\"u}tz, Philipp},
  title   = {A new database on the evolution of air--water flows along a large vertical pipe},
  journal = {International Journal of Thermal Sciences},
  volume  = {49},
  number  = {4},
  pages   = {664--674},
  year    = {2010},
  doi     = {10.1016/j.ijthermalsci.2009.11.006}
}

@article{Mudde2009,
  author  = {Mudde, Robert F. and Harteveld, Wim K. and van den Akker, Herman E. A.},
  title   = {Uniform Flow in Bubble Columns},
  journal = {Industrial \& Engineering Chemistry Research},
  volume  = {48},
  number  = {1},
  pages   = {148--158},
  year    = {2009},
  doi     = {10.1021/ie800630v}
}

@article{Hosokawa2009,
  author  = {Hosokawa, Satoru and Tomiyama, Akio},
  title   = {Multi-fluid simulation of turbulent bubbly pipe flows},
  journal = {Chemical Engineering Science},
  volume  = {64},
  number  = {24},
  pages   = {5308--5318},
  year    = {2009},
  doi     = {10.1016/j.ces.2009.08.018}
}

@article{Shawkat2008,
  author  = {Shawkat, Mohamed E. and Ching, Chan Y. and Shoukri, Mamdouh},
  title   = {Bubble and liquid turbulence characteristics of bubbly flow in a large diameter vertical pipe},
  journal = {International Journal of Multiphase Flow},
  volume  = {34},
  number  = {8},
  pages   = {767--785},
  year    = {2008},
  doi     = {10.1016/j.ijmultiphaseflow.2008.02.002}
}

@article{Lucas2005,
  author  = {Lucas, Dirk and Krepper, Eckhard and Prasser, Horst-Michael},
  title   = {Development of co-current air--water flow in a vertical pipe},
  journal = {International Journal of Multiphase Flow},
  volume  = {31},
  number  = {12},
  pages   = {1304--1328},
  year    = {2005},
  doi     = {10.1016/j.ijmultiphaseflow.2005.07.002}
}

@article{Hibiki2001,
  author  = {Hibiki, Takashi and Ishii, Mamoru and Xiao, Zheng},
  title   = {Axial interfacial area transport of vertical bubbly flows},
  journal = {International Journal of Heat and Mass Transfer},
  volume  = {44},
  number  = {10},
  pages   = {1869--1888},
  year    = {2001},
  doi     = {10.1016/S0017-9310(00)00217-5}
}

@article{Deen2001,
  author  = {Deen, Niels G. and Solberg, Terje and Hjertager, Bj{\o}rn H.},
  title   = {Large eddy simulation of the gas--liquid flow in a square cross-sectioned bubble column},
  journal = {Chemical Engineering Science},
  volume  = {56},
  number  = {21-22},
  pages   = {6341--6349},
  year    = {2001},
  doi     = {10.1016/S0009-2509(01)00255-4}
}

@article{Pfleger1999,
  author  = {Pfleger, Dietmar and Gomes, Susana and Gilbert, Nicolas and Wagner, Hans G.},
  title   = {Hydrodynamic simulations of laboratory scale bubble columns: Fundamental studies of the Eulerian--Eulerian modelling approach},
  journal = {Chemical Engineering Science},
  volume  = {54},
  number  = {21},
  pages   = {5091--5099},
  year    = {1999},
  doi     = {10.1016/S0009-2509(99)00279-2}
}

@inproceedings{Liu1998,
  author    = {Liu, Tong-Jen},
  title     = {The Role of Bubble Size on Liquid Phase Turbulent Structure in Two-Phase Bubbly Flow},
  booktitle = {Proceedings of the 3rd International Conference on Multiphase Flow (ICMF'98)},
  address   = {Lyon, France},
  dates     = {June 8--12},
  year      = {1998}
}

@article{LiuBankoff1993,
  author  = {Liu, Tong-Jen and Bankoff, S. George},
  title   = {Structure of air--water bubbly flow in a vertical pipe. II: Void fraction, bubble velocity and bubble size distribution},
  journal = {International Journal of Heat and Mass Transfer},
  volume  = {36},
  number  = {4},
  pages   = {1061--1072},
  year    = {1993},
  doi     = {10.1016/0017-9310(93)80029-8}
}

@article{Wang1987,
  author  = {Wang, S. K. and Lee, S. J. and Jones, O. C., Jr. and Lahey, R. T., Jr.},
  title   = {3-D turbulence structure and phase distribution measurements in bubbly two-phase flows},
  journal = {International Journal of Multiphase Flow},
  volume  = {13},
  number  = {3},
  pages   = {327--343},
  year    = {1987},
  doi     = {10.1016/0301-9322(87)90024-0}
}

@article{SunFaeth1986,
  author  = {Sun, T.-Y. and Faeth, G. M.},
  title   = {Structure of turbulent bubbly jets—II},
  journal = {International Journal of Multiphase Flow},
  volume  = {12},
  number  = {1},
  pages   = {115--126},
  year    = {1986},
  doi     = {10.1016/0301-9322(86)90010-4}
}

\end{document}